\documentclass{bioinfo}
\usepackage{hhline}
\usepackage{fancyhdr}
\usepackage[normalem]{ulem}
\usepackage{tabularx}
\usepackage{graphicx}

\usepackage{pdfpages}
\usepackage[LY1]{fontenc}

\usepackage{soul}

\usepackage{atbegshi,picture}
\usepackage{lipsum}
\usepackage{mathtools}
\usepackage{amsmath}

\usepackage{tikz}

\usepackage{floatrow}
\copyrightyear{2019} \pubyear{2019}

\access{Advance Access Publication Date: Day Month Year}
\appnotes{Manuscript Category}

\newcommand{\alser}[1]{{\color{black}{#1}}}

\begin{document}
\firstpage{1}

\subtitle{Subject Section}

\title[SneakySnake: A Fast and Accurate Universal Genome Pre-Alignment Filter]{\alser{SneakySnake: A Fast and Accurate Universal Genome Pre-Alignment Filter for CPUs, GPUs, and FPGAs}}

\author[Alser \textit{et~al}.]{
Mohammed Alser\,$^{\text{\sfb 1}, \text{\sfb 2},\ast}$,
Taha Shahroodi\,$^{\text{\sfb 1}}$,
Juan G\'{o}mez-Luna\,$^{\text{\sfb 1}, \text{\sfb 2}}$,
\\Can Alkan\,$^{\text{\sfb 4},\ast}$, and \alser{Onur Mutlu\,$^{\text{\sfb 1}, \text{\sfb 2}, \text{\sfb 3}, \text{\sfb 4},\ast}$}

}
\address{
$^{\text{\sf 1}}$Department of Computer Science, ETH Zurich, Zurich 8006, Switzerland\\
$^{\text{\sf 2}}$Department of Information Technology and Electrical Engineering, ETH Zurich, Zurich 8006, Switzerland\\
$^{\text{\sf 3}}$Department of Electrical and Computer Engineering, Carnegie Mellon University, Pittsburgh 15213, PA, USA \\
$^{\text{\sf 4}}$Department of Computer Engineering, Bilkent University, Ankara 06800, Turkey \\
}

\corresp{$^\ast$To whom correspondence should be addressed.}

\history{Received on XXXXX; revised on XXXXX; accepted on XXXXX}

\editor{Associate Editor: XXXXXXX}

\corresp{$^\ast$To whom correspondence should be addressed.}

\history{Received on XXXXX; revised on XXXXX; accepted on XXXXX}

\editor{Associate Editor: XXXXXXX}

\abstract{\textbf{Motivation:} 
We introduce \emph{SneakySnake}, a highly parallel and highly accurate pre-alignment filter that remarkably reduces the need for computationally costly sequence alignment. The key idea of SneakySnake is to reduce the \emph{approximate string matching} (ASM) problem to \alser{the} \emph{single net routing} (SNR) problem in VLSI chip layout. 
In \alser{the} SNR problem, we are interested in finding the optimal path that connects two terminals with the least routing cost on a special grid layout that contains obstacles. 
\alser{The} SneakySnake algorithm quickly solves the SNR problem and uses the found optimal path to decide whether {or not} performing sequence alignment is \alser{necessary}. Reducing the ASM problem into SNR also makes SneakySnake efficient to implement {on} 
CPUs, GPUs, and FPGAs.\\
\textbf{Results:} SneakySnake significantly improves the accuracy of pre-alignment filtering by up to four orders of magnitude compared to the state-of-the-art pre-alignment filters, Shouji, GateKeeper, and SHD. 
{For} short sequences, SneakySnake accelerates Edlib (state-of-the-art implementation of Myers's bit-vector algorithm) and Parasail ({state-of-the-art} sequence aligner with {a} configurable scoring function), by up to 37.7$\times$ and 43.9$\times$ (>12$\times$ on average), respectively, {with} its CPU implementation, and by up to 413$\times$ and 689$\times$ (>400$\times$ on average), respectively, {with} {FPGA and GPU} acceleration.
{For} long sequences, {the CPU implementation of} SneakySnake accelerates Parasail and KSW2 (sequence aligner of minimap2) by up to 979$\times$ (276.9$\times$ on average) and 91.7$\times$ (31.7$\times$ on average), respectively.
As SneakySnake does not replace sequence alignment, {users can still obtain \emph{all} capabilities (e.g., configurable scoring {functions}) of the aligner of their choice, unlike existing acceleration efforts that sacrifice some aligner capabilities}.\\
\textbf{Availability:} https://github.com/CMU-SAFARI/SneakySnake\\
\textbf{Contact:} \href{alserm@inf.ethz.ch}{alserm@inf.ethz.ch}, \href{calkan@cs.bilkent.edu.tr}{calkan@cs.bilkent.edu.tr}, \href{omutlu@ethz.ch}{omutlu@ethz.ch}\\
\textbf{Supplementary information:} Supplementary data is available at \emph{Bioinformatics} online.
\\}
\maketitle

\section{Introduction} \label{sec:introduction}
One of the most fundamental computational steps in most genomic analyses is {\emph{sequence alignment}}~\citep{alser2020technology, senol2019nanopore}. 
This step is formulated as \alser{an} \emph{approximate string matching} (ASM) problem \citep{navarro2001guided} and \alser{it calculates: (1) \emph{edit distance} between two given sequences, (2) type of each edit (i.e., insertion, deletion, or substitution), and (3) location of each edit in one of the two given sequences.}
Edit distance is defined as the minimum number of edits needed to convert one sequence into the other \citep{levenshtein1966binary}. \alser{These edits result from both sequencing errors \citep{firtina2020apollo} and genetic variations \citep{10002015global}. }
Edits can have different weights, based on a user-defined \emph{scoring} function, to allow favoring one edit type over \alser{another} \citep{wang2011comparison}. 
Sequence alignment involves a \emph{backtracking step}, which calculates an ordered list of characters representing the location and type of each possible edit operation required to change one of the two given sequences into the other. 
As any two sequences can have several different arrangements of the edit operations, we need to examine all possible \emph{prefixes} of the two input sequences and keep track of the pairs of prefixes that provide a minimum edit distance. 
Therefore, sequence alignment approaches are typically implemented as dynamic programming (DP) algorithms to avoid re-examining the same prefixes {many times} \citep{alser2020technology,eddy2004dynamic}. 
DP-based \alser{sequence alignment algorithms}, such as Needleman-Wunsch \citep{needleman1970general}, are computationally expensive as they have quadratic time and space complexity (i.e., O($m^2$) for a sequence length of $m$). Many attempts were made to boost the performance of existing sequence aligners. Recent attempts tend to follow one of two \alser{key directions}{, as we comprehensively survey} in~\citep{alser2020accelerating}: (1) Accelerating the DP algorithms using hardware accelerators and (2) Developing pre-alignment filtering heuristics that reduce the need for the DP algorithms, given an edit distance threshold.

Hardware accelerators include building aligners that use 1) multi-core and SIMD (single instruction multiple data) capable central processing units (CPUs), such as Parasail \citep{daily2016parasail}.
The classical DP algorithms can also be accelerated by calculating a bit representation of the DP matrix and processing its bit-vectors in parallel, such as Myers's bit-vector algorithm \citep{myers1999fast}. To our knowledge, Edlib \citep{vsovsic2017edlib} is currently the best-performing implementation of Myers's bit-vector algorithm. 
Other hardware accelerators include 2) graphics processing units (GPUs), such as GSWABE \citep{liu2015gswabe}, 3) field-programmable gate arrays (FPGAs), such as FPGASW \citep{fei2018fpgasw}, or 4) {processing-in-memory architectures that enable performing computations inside the memory chip and alleviate the need for transferring the data to the CPU cores, such as} GenASM~\citep{senolcalimicro2020}. However, many of these efforts either simplify the scoring function as in Edlib, or only take into account accelerating the computation of the DP matrix without performing the backtracking step as in \citep{ chen2014accelerating}. Different and more sophisticated scoring functions are typically needed to better quantify the similarity between two sequences \citep{wang2011comparison}. The backtracking step involves unpredictable and irregular memory access patterns, which pose a difficult challenge for efficient hardware implementation.

Pre-alignment filtering heuristics aim to quickly eliminate some of the dissimilar sequences before using the computationally-expensive optimal alignment algorithms.
Existing pre-alignment filtering techniques are either: 1) slow and they suffer from a limited sequence length ($\leq128 bp$), such as SHD \citep{xin2015shifted}, or 2) inaccurate after some edit distance threshold, such as GateKeeper \citep{alser2017gatekeeper} and MAGNET \citep{alser2017magnet1}.
{Highly-parallel filtering can also be achieved using processing-in-memory architectures, as in} GRIM-Filter~\citep{kim2018grim}.
Shouji \citep{alser2019shouji} is currently the best-performing FPGA pre-alignment filter in terms of both accuracy and execution time.

Our \textbf{goal} in this work is to significantly reduce the time spent on calculating the sequence alignment of \emph{both short and long sequences} using very fast and accurate pre-alignment filtering. 
To this end, we introduce \emph{SneakySnake}, a highly parallel and highly accurate pre-alignment filter that works on {\emph{modern} high-performance computing architectures such as CPUs, GPUs, and FPGAs}. \alser{The \textbf{key idea} of SneakySnake is to provide {a} highly-accurate pre-alignment filtering algorithm by reducing the ASM problem to the \emph{single net routing} (SNR) problem \citep{lee1976use}. 
The SNR problem is to find the shortest routing path that interconnects two terminals on the boundaries of VLSI chip layout {while passing} through the minimum number of obstacles.
Solving the SNR problem is faster than solving the ASM problem, as calculating the routing path after facing an obstacle is independent of the calculated path before this obstacle.}
This provides two key benefits. 1) It obviates the need for using computationally costly DP algorithms to keep track of the subpath that provides {the} optimal {solution} (i.e., {the one} with the least possible routing cost).
2) The independence {of} the subpaths allows for solving many SNR subproblems in parallel by judiciously leveraging the parallelism-friendly architecture of modern FPGAs and GPUs to greatly speed up the SneakySnake algorithm.

The \textbf{contributions} of this paper are as follows: 
\begin{itemize}
\item We introduce SneakySnake, 
the fastest and most accurate pre-alignment filtering mechanism to date that greatly enables the speeding up of genome sequence alignment while preserving its accuracy. We demonstrate that the SneakySnake algorithm is 1) correct and optimal in solving the SNR problem and 2) it runs in linear time with respect to sequence length and edit distance threshold.
\item We demonstrate that the SneakySnake algorithm significantly improves the accuracy of pre-alignment filtering by up to four orders of magnitude compared to Shouji, GateKeeper, and SHD.
\item We provide, to our knowledge, the \emph{first universal} pre-alignment filter for {CPUs, GPUs, and FPGAs}, by having software as well as software/hardware co-designed versions {of SneakySnake}.
\item We demonstrate, using short sequences, that SneakySnake accelerates Edlib and Parasail by up to 37.7$\times$ and 43.9$\times$ (>12$\times$ on average), respectively, {with} its CPU implementation, and by up to 413$\times$ and 689$\times$ (>400$\times$ on average), respectively, {with} {FPGA and GPU} acceleration. 
{We also demonstrate, using long sequences, that SneakySnake accelerates Parasail by up to 979$\times$ (276.9$\times$ on average).}
\item We demonstrate that {the CPU implementation of} SneakySnake accelerates the sequence alignment of minimap2 \citep{li2018minimap2}, a state-of-the-art read mapper, {by up to 6.83$\times$ and  91.7$\times$ using short and long sequences, respectively}.
\end{itemize}
\section{Methods} \label{sec:methods}

\subsection{Overview}
\alser{The primary purpose of SneakySnake is to accelerate sequence alignment calculation by providing fast and accurate pre-alignment filtering}. 
The SneakySnake algorithm quickly examines each sequence pair before applying sequence alignment and decide{s} whether computationally{-}expensive sequence alignment is needed \alser{for two genomic sequences}. 
This filtering decision of the SneakySnake algorithm is made based on accurately estimating the number of edits between two given sequences.  
If two genomic sequences differ by more than the edit distance threshold, then the two sequences are identified as dissimilar sequences and hence identifying the location and the type of each edit is not needed. 
\emph{{The edit distance estimated by the SneakySnake algorithm should always be less than or equal to the actual edit distance}} value so that SneakySnake ensures \emph{reliable and lossless} filtering (preserving all similar sequences).
\alser{To reliably estimate the edit distance between two sequences, we reduce the ASM problem to the SNR problem. 
That is, instead of calculating the sequence alignment, the SneakySnake algorithm finds the routing path that interconnects two terminals {while passing} through the minimum number of obstacles on {a} VLSI chip.
The number of obstacles faced throughout the found routing path represents a \emph{lower bound} on the edit distance between two sequences (Theorem 2, Section \ref{theorem}) and hence this number of obstacles can be used for the reliable filtering decision of SneakySnake. 
SneakySnake treats all obstacles (edits) faced along a path equally (i.e., it does not favor one type of edits over the others).
This eliminates the need for examining different possible arrangements of the edit operations, as in DP-based algorithms, and makes solving the SNR problem easier and faster than solving the ASM problem. However, users can still configure the aligner of their choice for their desired scoring function.} 


\subsection{Single Net Routing (SNR) Problem} \label{SNR_problem}
The SNR problem in VLSI chip layout \alser{refers to the problem of optimally interconnecting two terminals on a grid graph while respecting constraints}. We present an example of a VLSI chip layout in Fig. \ref{fig:overview}. The goal is to find the optimal path \alser{--called \emph{signal net}--} that connects the source and destination terminals through the chip layout. We describe the special grid graph of the SNR problem and define {the} optimal signal net as follows:

\begin{figure}
\centerline{\includegraphics[width=1\linewidth]{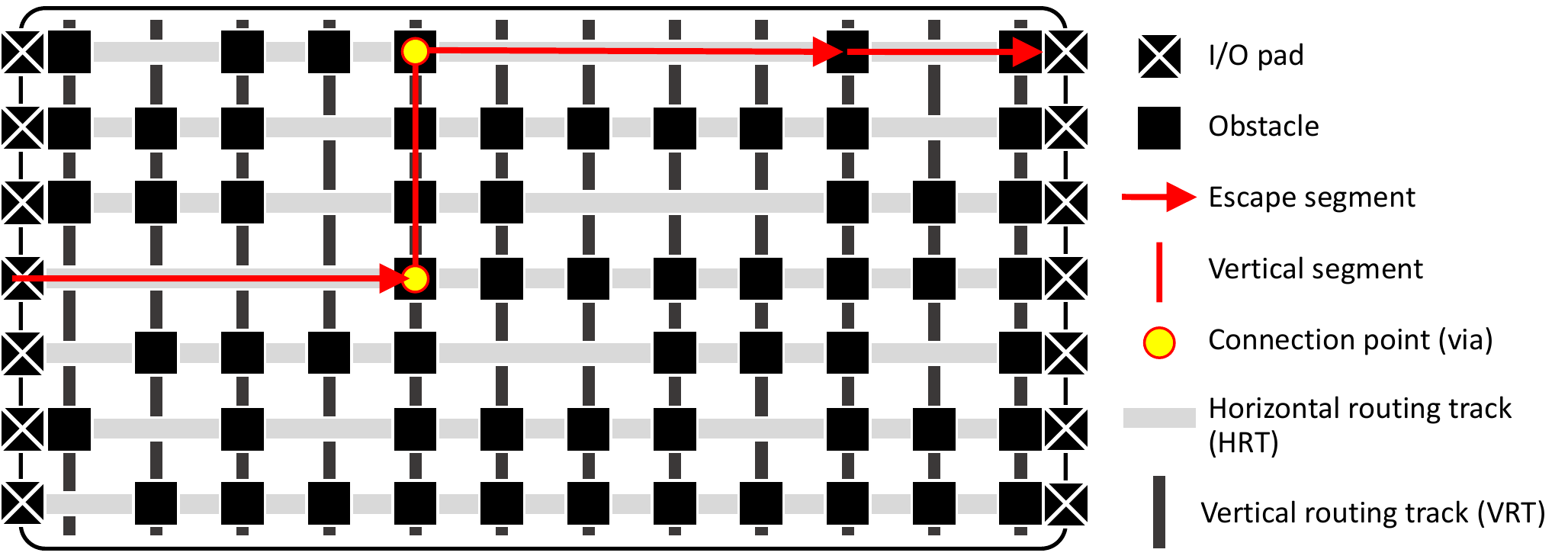}}
\caption{\alser{{Chip layout with processing elements and two layers of metal routing tracks. In this example, the chip layout has 7 horizontal routing tracks (HRTs) located on the first layer and another 12 vertical routing tracks (VRTs) located on the second layer. 
The optimal signal net that is calculated using the SneakySnake algorithm 
is highlighted in red using three escape segments. The first escape segment is connected to the second escape segment using {a} VRT through vias. The second escape segment is connected to the third escape segment without passing through a VRT as both escape segments are located on the same HRT. The optimal signal net passes through three {obstacles (each of which is located at the end of each escape segment)} and hence the signal net has a total delay of} $3\times t_{obstacle}$.}}
\label{fig:overview}
\end{figure}

\begin{itemize}

\item The chip layout has two layers of evenly spaced metal routing tracks. While the first layer allows traversing the chip horizontally through dedicated \emph{horizontal routing tracks} (HRTs), the second layer allows traversing the chip vertically using dedicated \emph{vertical routing tracks} (VRTs).

\item The horizontal and vertical routing tracks induce a two dimensional uniform grid over the chip layout. Each HRT can be obstructed by some obstacles (e.g., processing elements in the chip). \alser{For simplicity, we assume that VRTs can not be obstructed by obstacles.} These obstacles allow the signal to pass horizontally through HRTs, but they induce a signal delay on the passed signal. Each obstacle induces a fixed propagation delay, $t_{obstacle}$, on the victim signal that passes through the obstacle in the corresponding HRT.
\item A signal net often uses a sequence of alternating horizontal and vertical segments that are parts of the routing tracks. Adjacent horizontal and vertical segments in the signal net are connected by an inter-layer \emph{via}. \alser{We call a signal net \emph{optimal} if it is both the shortest and the fastest routing path 
(i.e., passes through the minimum number of obstacles)}.
\item \alser{Alternating between horizontal and vertical segments is restricted by passing a single obstacle. Thus, segment alternating strictly} delays the signal by $t_{obstacle}$ time.
\item The terminals can be any of the I/O pads that are located on the right-hand and left-hand boundaries of the chip layout. The source terminal always lies on the opposite side of the destination terminal.
\end{itemize}

The general goal of this SNR problem is to find an \emph{optimal} signal net in the grid graph of the chip layout. 
For the simplicity of developing a solution, we call a horizontal segment that ends with at most an obstacle an \emph{escape segment}. 
The escape segment can also be a single obstacle only. Also for simplicity, we call the right-hand side of an escape segment {a} \emph{checkpoint}. 
Next, we present how we can reduce the ASM problem to the SNR problem.

\subsection{Reducing the Approximate String Matching (ASM) Problem to the Single Net Routing (SNR) Problem}\label{ASM2SNR}
We reduce the problem of finding the similarities and differences between two genomic sequences to that of finding the optimal signal net in a VLSI chip layout.
\alser{Reducing the ASM problem to the SNR problem requires two key steps: (1) replacing the DP table used by \alser{the} sequence alignment algorithm \alser{to} a special grid graph called \emph{chip maze} and (2) finding the number of differences between two genomic sequences} \unskip\parfillskip 0pt \par

\begin{figure*}
\centerline{\includegraphics[width=0.95\linewidth]{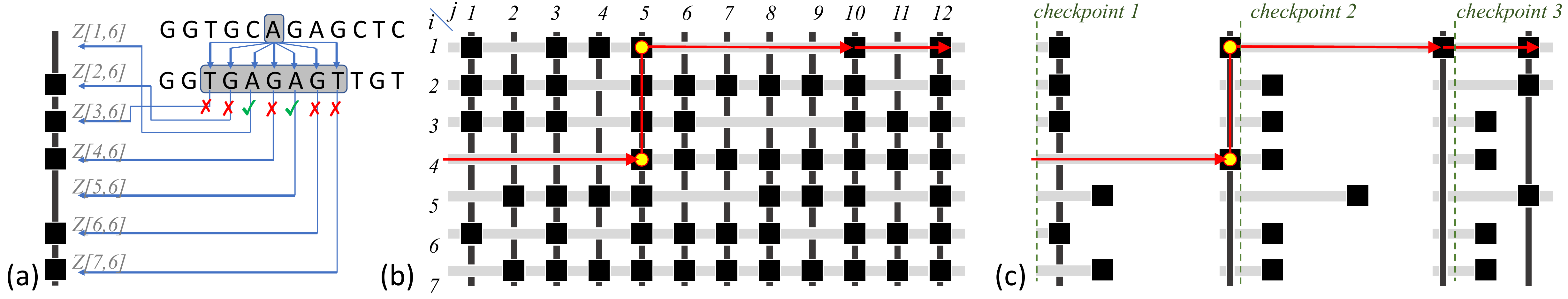}}
\caption{{(a) An example of how we build the 6$^{th}$ column of the chip maze, $Z$, using Equation 1 for \alser{a reference sequence} $R$ = \lq{GGTGCAGAGCTC}\rq{}, \alser{a query sequence} $Q$ = \lq{GGTGAGAGTTGT}\rq{}, and an edit distance threshold ($E$) of 3. The 6$^{th}$ character of $R$ is compared with each of its corresponding $2E+1$ characters of $Q$. 
{The order of the results of comparing $R[6]$ with $Q[3]$, $Q[4]$, and $Q[5]$ is reversed to easily derive the second case of Equation 1}.
(b) The complete chip maze that is calculated using Equation 1, which has 2$E$+1 rows and $m$ (length of $Q$) columns. (c) The actual chip maze that is calculated using the SneakySnake algorithm. The optimal signal net is highlighted in both chip mazes in red. The signal net has 3 obstacles {(each of which is located at the end of each escape segment)} and hence sequence alignment is needed, as the number of differences $\leq E$.}}
\label{fig:steps}
\end{figure*}

\noindent \alser{in the chip maze by solving the SNR problem.} 
We replace the $(m+1) \times (m+1)$ DP table with our chip maze, $Z$, where $m$ is the sequence length (for simplicity, we assume that we have a pair of equal-length sequences but we relax this assumption in {Section}~\ref{theorem}). 
The chip maze is a $(2E+1) \times m$ grid graph, where $E$ is the edit distance threshold {in terms of {the number of} tolerable character differences}, $(2E+1)$ is the number of HRTs, and $m$ is the number of VRTs. 
The chip maze is an abstract layout for the VLSI chip layout, \alser{as we show in Fig. \ref{fig:steps}(b) for the same chip layout of Fig. \ref{fig:overview}}. 
Each entry of the chip maze represents the pairwise comparison result of a character of one sequence with another character of the other sequence. 
A pairwise mismatch is represented by an obstacle (an entry of value '1') in the chip maze and a pairwise match is represented by an available path (an entry of value '0') in its corresponding HRT.
Given two genomic sequences, a reference sequence $R[1\dots m]$ and a query sequence $Q[1\dots m]$, and an edit distance threshold $E$, we calculate the entry $Z[i, j]$ of the chip maze, where $1\leq i\leq (2E+1)$ and $1\leq j\leq m$, as follows:\newline
\begin{equation}
\label{equ:chipmaze}
Z[i,j]=\left\{\begin{matrix*}[l]
0,\quad\quad if~~i=E+1, ~Q[j]=R[j],  \\ 
0,\quad\quad if~~1\leq i\leq E, ~Q[j-i]=R[j],  \\ 
0,\quad\quad if~~i> E+1, ~Q[j+i-E-1]=R[j],  \\ 
1,\quad\quad otherwise 
\end{matrix*}\right.
\end{equation}

We derive {the four cases of Equation 1} by considering  all possible pairwise matches and mismatches (due to possible edits) between two sequences. That is, each column of the chip maze stores the result of comparing the $j^{th}$ character of the \alser{reference sequence}, $R$, with each of {the} corresponding $2E+1$ characters of the \alser{query sequence}, $Q${, as we show in Fig.} \ref{fig:steps}(a). 
{In the first case of Equation 1, we compare the $j^{th}$ character of the reference sequence, $R$, with the $j^{th}$ character of the query sequence, $Q$, to detect pairwise matches and substitutions.
In the second case of Equation 1, we compare the $j^{th}$ character of the reference sequence with each of the $E$ left-hand neighboring characters of the $j^{th}$ character of the query sequence, to accurately detect deleted characters in the query {sequence}.
In the third case of Equation 1, we compare the $j^{th}$ character of the reference sequence with each of the $E$ right-hand neighboring characters of the $j^{th}$ character of the query sequence, to accurately detect inserted characters in the query {sequence}}. 
Each insertion and deletion can shift multiple trailing characters (e.g., deleting the character \lq{N}\rq{} from \lq{GE\textbf{N}OME}\rq{} shifts the last three characters to the left direction, making it \lq{GE\textbf{OME}}\rq{}). {Hence, in the second and the third cases of Equation 1,} we need to compare a character of the \alser{reference sequence} with the neighboring characters of its corresponding character of the \alser{query sequence} to cancel the effect of deletion/insertion and correctly detect the common subsequences between two sequences. 
{In the fourth case of Equation 1,} we fill the remaining empty entries of each row {with} ones (i.e., obstacles) to indicate that there is no match between the corresponding characters.
{These four cases are essential to accurately detect} substituted, deleted, and inserted characters in one or both {of the} sequences. 
We present in Fig. \ref{fig:steps}(b) an example of the chip maze for two sequences, where a \alser{query sequence}, $Q$, differs from a \alser{reference sequence}, $R$, by three edits. 

The chip maze is a data-dependency free data structure as computing each of its entries is independent of every other and thus the entire grid graph can be computed all at once in a parallel fashion. Hence, our chip maze is well suited for both sequential and highly-parallel computing platforms \citep{seshadri2017ambit}.
The challenge is now calculating the minimum number of edits between two sequences using the chip maze. 
Considering the chip maze as a chip layout where the rows represent the HRTs and the columns represent the VRTs, we observe that we can reduce the ASM problem to the SNR problem. 
Now, the problem becomes finding an optimal set (i.e., signal net) of non-overlapping escape segments. As we discuss in Section \ref{SNR_problem}, a set of escape segments is optimal if there is no other set that solves the SNR problem and has both {smaller} number of escape segments and {smaller} number of entries of value '1' (i.e., obstacles). Once we find such an optimal set \alser{of escape segments}, we can compute the minimum number of edits between two sequences as the total number of obstacles along the computed optimal set. 
Next, we present an efficient algorithm that solves this SNR problem.



\subsection{Solving the Single Net Routing Problem}\label{theorem}
\alser{The primary purpose of the SneakySnake algorithm is to solve the SNR problem by providing an optimal signal net. 
Solving the SNR problem requires achieving two key objectives: 1) achieving the lowest possible latency by finding the minimum number of escape segments that are sufficient to link the source terminal to the destination terminal and 2) achieving the shortest length of the signal net by considering each escape segment just once and in monotonically increasing order of their start index (or end index). 
The first objective is based on a key observation that a signal net with fewer escape segments always {has} fewer obstacles, as each escape segment has at most a single obstacle (based on our definition in Section \ref{SNR_problem}). This key observation leads to a signal net that has the least possible total propagation delay. The second objective restricts the SneakySnake algorithm from ever searching backward for the longest escape segment. This leads to a signal net that has non-overlapping escape segments.

To achieve these two key objectives, the SneakySnake algorithm applies five effective steps. (1) The SneakySnake algorithm first constructs the chip maze {using Equation 1}. 
It then considers the first column of the chip maze as the first checkpoint, where the first iteration starts.
(2) At each new checkpoint, the SneakySnake algorithm always selects the longest escape segment that allows the signal to travel as far forward as possible until it reaches an obstacle. 
For each row of the chip maze, it computes the length of the first horizontal segment of consecutive entries of value '0' that starts from a checkpoint and ends at an obstacle or at the end of the current row. 
The SneakySnake algorithm compares the length of all the $2E+1$ computed horizontal segments, selects the longest one, and considers it along with its first following obstacle as an escape segment. 
If the SneakySnake algorithm is unable to find a horizontal segment (i.e., following a checkpoint, all rows start with an obstacle), it considers one of the obstacles as the longest escape segment. 
It considers the computed escape segment as part of the solution to the SNR problem. 
(3) It creates a new checkpoint after the longest escape segment. 
(4) It repeats the second and third steps until either the signal net reaches a destination terminal, or the total propagation delay exceeds the allowed propagation delay threshold (i.e., $E\times t_{obstacle}$).} 
{When the two input sequences are different in length, {we} need to count the number of obstacles more conservatively along {the signal net}. 
Doing so ensures a correct reduction of the ASM problem. 
This means that we need to deduct the total number of leading and trailing obstacles from the total count of edits between two input sequences before making the filtering decision, as such obstacles can be caused by the fourth case of Equation 1.} 
(5) If SneakySnake finds the optimal net using the previous steps, then {it indicates that the edit distance between two input sequences is $\leq E$}.
{If so, sequence alignment is needed to know the exact number of edits, type of each edit, and location of each edit between the two} sequences using user's favourite sequence alignment algorithm. 
Otherwise, the SneakySnake algorithm terminates without performing computationally expensive sequence alignment{, since the differences between sequences is guaranteed to be $>E$}.

To efficiently implement the SneakySnake algorithm, we use an implicit representation of the chip maze. That is, the SneakySnake algorithm starts computing on-the-fly one entry of the chip maze after another for each row until it faces an obstacle (i.e., $Z$[$i$,$j$] = 1) or it reaches the end of the current row. 
Thus, the entries that are actually calculated for each row of the chip maze are the entries that are located only between each checkpoint and the first obstacle, in each row, following this checkpoint, as we show in Fig. \ref{fig:steps}(c). 
This significantly reduces the number of computations needed for the SneakySnake algorithm.
We provide the SneakySnake algorithm along with analysis of its computational complexity (asymptotic run time and space complexity) in \textcolor{black}{Supplementary Materials, Section 5}.

The SneakySnake algorithm is both correct and optimal {in solving the SNR problem}. 
The SneakySnake algorithm is correct as it always provides a signal net (if it exists) that interconnects the source terminal and the destination terminal. 
In other words, it does not lead to routing failure as signal will eventually reach its destination.

\textbf{Theorem 1.} \emph{The SneakySnake algorithm {is guaranteed} to find a signal net that interconnects the source terminal and the destination terminal when one exists.\\}
We provide the correctness proof for Theorem 1 in \textcolor{black}{Supplementary Materials, Section 6.1}. 
The SneakySnake algorithm is also optimal as it {is guaranteed} to find an optimal signal net that links the source terminal to destination terminal when one exists. 
Such an optimal signal net always ensures that the signal arrives the destination terminal with the least possible total propagation delay.

\textbf{Theorem 2.} \emph{When a signal net exists between the source terminal and the destination terminal, using the SneakySnake algorithm, a signal from the source terminal reaches the destination terminal with the minimum possible latency.\\}
We provide the optimality proof for Theorem 2 in \textcolor{black}{Supplementary Materials, Section 6.2}.
\\\vspace{-10 pt}

{\textbf{Different from existing sequence alignment algorithms} that are based on DP approaches} \citep{daily2016parasail, xin2013accelerating} or sparse DP (i.e., chaining exact matches between two sequences using DP algorithms) approaches \citep{chaisson2012mapping}, 
{SneakySnake 1) does not require knowing the location and the length of {common subsequences} between the two {input} sequences in advance, 2) does not consider the vertical distance (i.e., the number of rows) between two escape segments in the calculation of the minimum number of edits, and 3) does not build the entire dynamic programming table; SneakySnake builds only a minimal portion of the chip maze that is needed to provide an optimal solution.
The first difference makes SneakySnake independent of any algorithm that aims to calculate sequence alignment, as SneakySnake quickly and efficiently calculates its own data structure (i.e., chip maze) to find \emph{all} {common subsequences}. {The second difference helps {to construct} a data dependency-free chip maze and allows for solving many SNR subproblems in parallel as calculating the routing path after facing an obstacle is independent of the calculated path before this obstacle.}
The third difference significantly reduces the number of computations needed for the SneakySnake algorithm.}


\textbf{Different from existing edit distance approximation algorithms} \citep{chakraborty2018approximating,charikar2018estimating}
that sacrifice the optimality of the edit distance solution (i.e., its solution $\geq$ the actual edit distance of each sequence pair) for a reduction in time complexity, (e.g., $O(m^{1.647})$ instead of $O(m^{2})$), SneakySnake does not overestimate the edit distance as the calculated optimal signal net has \emph{always} the minimum possible number of obstacles (Theorem 2).
{We {take advantage of} the edit distance underestimation of SneakySnake by using our fast computation method as a pre-alignment filter}.
{Doing so} {ensures} two key properties: \alser{(1) {allows} sequence alignment to be calculated only for similar (or nearly similar) sequences and (2) {accelerates} the sequence alignment algorithms without changing (or replacing) their algorithmic method and hence preserving all the capabilities of the sequence alignment algorithms}.

We next discuss further optimizations and new software/hardware co-designed versions of the SneakySnake algorithm that {can} leverage FPGA and GPU architectures for highly-parallel computation.\\ 

\subsection{Snake-on-Chip Hardware Architecture} \label{Snake-on-Chip-Section}
We introduce an FPGA-friendly architecture for the SneakySnake algorithm, called \emph{Snake-on-Chip}. The main idea behind the hardware architecture of Snake-on-Chip is to divide the SNR problem into smaller non-overlapping subproblems. Each subproblem has a width of $t$ VRTs and a height of $2E+1$ HRTs, where $1<t\leq m$. We then solve each subproblem independently from the other subproblems. This approach results in three key benefits. 
(1) Downsizing the search space into a reasonably small grid graph with a known dimension {at} design time limits the number of all possible solutions for that subproblem. This reduces the size of the look-up tables (LUTs) required to build the architecture and simplifies the overall design. 
(2) Dividing the SNR problem into subproblems helps to maintain a modular and scalable architecture that can be implemented for any sequence length and edit distance threshold. (3) All the smaller subproblems can be solved independently and rapidly {with} high parallelism. This reduces the execution time of the overall algorithm as the SneakySnake algorithm does not need to evaluate the entire chip maze. 

However, these three key benefits come at the cost of accuracy degradation. As we demonstrate in Theorem 2, the SneakySnake algorithm guarantees to find an optimal solution to the SNR problem. However, the solution for each subproblem is not necessarily part of the optimal solution for the main problem (with the original size of $(2E+1) \times m$). This is because the source and destination terminals of these subproblems are not necessarily the same. 
{The SneakySnake algorithm determines the source and destination terminals for each SNR subproblem based on the optimal signal net of each SNR subproblem.}
This {leads to underestimation of} the total number of obstacles found along each signal net of each SNR subproblem. This is still acceptable as long as {the SneakySnake algorithm} solves the SNR problem quickly and {\emph{without}} {\emph{overestimating}} the number of obstacles {compared to the edit distance threshold}.
We provide the details of our hardware architecture of Snake-on-Chip in \textcolor{black}{Supplementary Materials, Section 8}.

\subsection{Snake-on-GPU Parallel Implementation} \label{Snake-on-GPU-Section}
{We} introduce our GPU implementation of the SneakySnake algorithm, called \emph{Snake-on-GPU}.
The main idea of Snake-on-GPU is to exploit the large number (typically few thousands) of GPU threads provided by modern GPUs to solve a large number of SNR problems rapidly and concurrently. 
In Snake-on-Chip, we explicitly divide the SNR problem into smaller non-overlapping subproblems and then solve all subproblems concurrently and independently using our specialized hardware. 
In Snake-on-GPU, we follow a different approach than that of Snake-on-Chip by keeping the same size of the original SNR problem and solving a massive number of these SNR problems at the same time. 
Snake-on-GPU uses one single GPU thread to solve one SNR problem (i.e., comparing one query sequence to one reference sequence at a time). 
This granularity of computation fits well the amount of resources (e.g., registers) that are available to each GPU thread and avoids the need for synchronizing several threads working on the same SNR problem. 

Given the large size of the sequence pair dataset that the GPU threads need to access, we carefully design Snake-on-GPU to efficiently 1) copy the input dataset of query and reference sequences into the GPU global memory, which is the off-chip DRAM memory of GPUs~\citep{cuda-guide} and it typically fits a few GB of data and 2) allow each thread to store its own query and reference sequences using the on-chip register file to avoid unnecessary accesses to the off-chip global memory. Each thread solves the complete SNR problem for a single query sequence and a single reference  sequence.
We provide the details of our parallel implementation of Snake-on-GPU in \textcolor{black}{Supplementary Materials, Section 9}.

\section{Results} \label{sec:results}
{We} evaluate 1) filtering accuracy, 2) filtering time, and 3) benefits of combining our universal implementation of the SneakySnake algorithm with state-of-the-art aligners.
We provide {a comprehensive treatment {of} all evaluation results in the {Supplementary Excel File} and {on the SneakySnake GitHub page}.
We compare the performance of SneakySnake, Snake-on-Chip, and Snake-on-GPU to {four} pre-alignment filters, Shouji~\citep{alser2019shouji}, MAGNET~\citep{alser2017magnet1}, GateKeeper~\citep{alser2017gatekeeper}, and SHD~\citep{xin2015shifted}. 
{We run the experiments that use multithreading and long sequences {on} a 2.3 GHz Intel Xeon Gold 5118 CPU with up to 48 threads and 192 GB RAM.} 
We run all other experiments {on} a 3.3 GHz Intel E3-1225 CPU with 32 GB RAM. We use a Xilinx Virtex 7 VC709 board \citep{guide2013virtex} to implement Snake-on-Chip and other existing accelerator architectures {(}Shouji, MAGNET, and GateKeeper). We build the FPGA design using Vivado 2015.4 in synthesizable Verilog. We use {an} NVIDIA GeForce RTX 2080Ti card \citep{guide2019Nvidia2080} with a global memory of 11 GB {GDDR6} to implement Snake-on-GPU. Both Snake-on-Chip and Snake-on-GPU are \emph{independent} of the specific FPGA and GPU platforms as they do not rely on any vendor-specific computing elements (e.g., intellectual property cores).

\subsection{{Evaluated Datasets}} \label{sec:experiment}
Our experimental evaluation uses 4 different real datasets (\texttt{100bp\_1}, \texttt{100bp\_2}, \texttt{250bp\_1}, and \texttt{250bp\_2}) and 2 simulated datasets (\texttt{10Kbp} and \texttt{100Kbp}). 
Each real dataset contains 30 million real sequence pairs (text and query pairs). \texttt{100bp\_1} and \texttt{100bp\_2} have sequences of length 100 bp, while \texttt{250bp\_1} and \texttt{250bp\_2} have sequences of length 250 bp.
We generate {the} \texttt{10Kbp} dataset to have 100,000 sequence pairs, each of which is 10 Kbp long, while {the} \texttt{100Kbp} dataset has 74,687 sequence pairs, each of which is 100 Kbp long.
{Supplementary Materials, Section 10.1 provides the details of these datasets}.

\subsection{Filtering Accuracy}\label{filtering_accuracy}
We evaluate the accuracy of {a} pre-alignment filter by computing its rate of falsely-accepted and falsely-rejected sequences before performing sequence alignment. The false accept rate is the ratio of the number of dissimilar sequences that are falsely accepted by the filter and the number of dissimilar sequences that are rejected by the sequence alignment algorithm. 
The false reject rate is the ratio of the number of similar sequences that are rejected by the filter and the number of similar sequences that are accepted by the sequence alignment algorithm. A reliable pre-alignment filter should always ensure both a 0\% false reject rate {to maintain the correctness of {the} genome analysis pipeline and an \emph{as-small-as-possible}} false accept rate  {to} maximize the number of dissimilar sequences that are eliminated {at low performance overhead}.

We first assess the false accept rate of SneakySnake, Shouji, MAGNET, GateKeeper, and SHD across different four real datasets and edit distance thresholds of $0\%-10\%$ of the sequence length. In Fig. \ref{fig:false-accept-rate}, 
we provide the false accept rate of each of the five filters.
We use Edlib to identify the ground{-}truth truly-accepted sequences for each edit distance threshold. Based on Fig. \ref{fig:false-accept-rate}, we make four key observations. 
(1) SneakySnake provides the lowest false accept rate compared to all the four state-of-the-art pre-alignment filters. 
SneakySnake provides up to 31412$\times$, 20603$\times$, and 64.1$\times$ less number of falsely-accepted sequences compared to GateKeeper/SHD (using \texttt{250bp\_2}, $E$= 10\%), Shouji (using \texttt{250bp\_2}, $E$= 10\%), and MAGNET (using \texttt{100bp\_1}, $E$= 1\%), respectively.
(2) MAGNET provides the second lowest false accept rate. It provides up to 25552$\times$ and 16760$\times$ less number of falsely-accepted sequences compared to GateKeeper/SHD (using \texttt{250bp\_2}, $E$= 10\%) and Shouji (using \texttt{250bp\_2}, $E$= 10\%), respectively.
(3) All five pre-alignment filters are less accurate in examining \texttt{100bp\_1} and \texttt{250bp\_1} than the other datasets, \texttt{100bp\_2} and \texttt{250bp\_2}. {This is expected as the actual number of edits of most of the sequence pairs in \texttt{100bp\_1} and \texttt{250bp\_1} datasets is very close to the edit distance threshold (Supplementary Materials, Table 4) and hence any underestimation in calculating the edit distance can lead to {falsely{-}accepted sequence pairs (i.e.,} estimated edit distance} $\leq E$).
(4) GateKeeper and SHD become ineffective for edit distance thresholds of greater than 8\% and 3\% for sequence lengths of 100 and 250 characters, respectively, as they accept all the input sequence pairs. 
This causes {a read mapper using} them to examine each sequence pair unnecessarily twice (i.e., once by GateKeeper or SHD and once by the sequence alignment algorithm).

Second, we {find that} SneakySnake {has} a 0\% false reject rate {(not plotted)}. {This observation is in accord with our theoretical proof of Theorem 2.}
It is also demonstrated in \citep{alser2019shouji} that Shouji and GateKeeper have a 0\% false reject rate, while MAGNET can falsely reject some similar sequence pairs.

We conclude that SneakySnake improves the accuracy of pre-alignment filtering by up to four orders of magnitude compared to the state-of-the-art pre-alignment filters. We also conclude that SneakySnake is the most effective pre-alignment filter, with a very low false accept rate and a 0\% false reject rate across a wide range of both edit distance thresholds and sequence lengths.

\begin{figure}
\centerline{\includegraphics[width=1\linewidth]{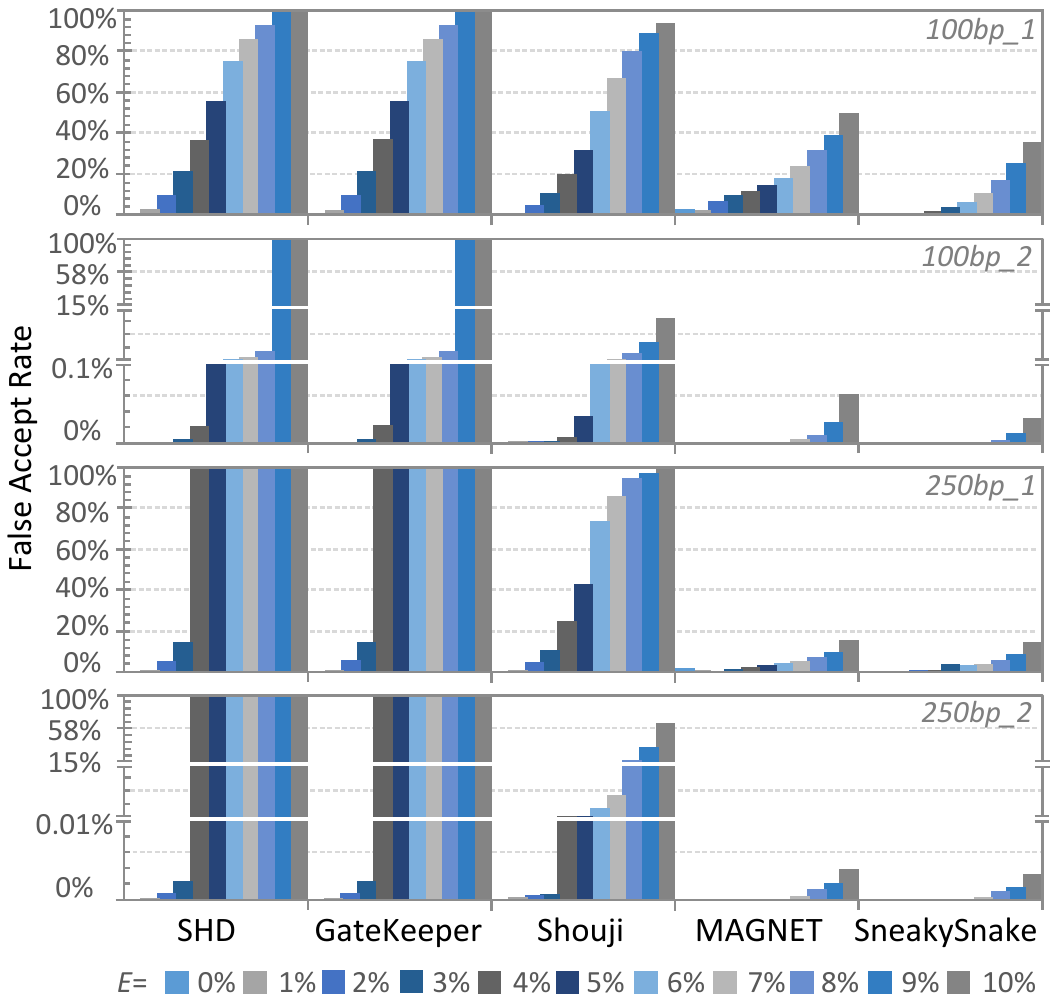}}
\caption{{False accept rates of SHD, GateKeeper, Shouji, MAGNET, and SneakySnake across 4 real datasets of short sequences}. We use a wide range of edit distance thresholds ($0\%-10\%$ of the sequence length) for sequence lengths of 100 and 250 bp.}
\label{fig:false-accept-rate}
\end{figure}

\subsection{{Effect} of SneakySnake on Short Sequence Alignment}
We analyze the benefits of integrating CPU-based pre-alignment filters, SneakySnake and SHD with the state-of-the-art CPU-based sequence aligners, Edlib and Parasail. We evaluate all tools using a single CPU core and single thread environment. Fig. \ref{fig:CPUSneakySnake}(a) and (b) present the normalized end-to-end execution time of SneakySnake and SHD{,} each combined with Edlib and Parasail, using our four real datasets over edit distance thresholds of $0\%-10\%$ of the sequence length. We make four key observations. (1) The addition of SneakySnake as a pre-alignment filtering step {significantly} reduces the execution time of Edlib and Parasail by up to 37.7$\times$ (using \texttt{250bp\_2}, $E$= 0\%) and 43.9$\times$ (using \texttt{250bp\_2}, $E$ =2\%), respectively. 
We also observe a similar trend as the number of CPU threads increases from 1 to 40, as we {show} in Supplementary Materials, Section 10.2.
{To explore the reason for this significant speedup, we need to check how fast SneakySnake examines the sequence pairs compared to sequence alignment, which we observe next}.
(2) SneakySnake is up to 43$\times$ (using \texttt{250bp\_1}, $E$= 0\%) and  47.2$\times$ (using \texttt{250bp\_1}, $E$= 2\%) faster than Edlib and Parasail, respectively, in examining the sequence pairs. 
(3) SneakySnake provides up to 8.9$\times$ 
and 40$\times$ 
more speedup to the end-to-end execution time of Edlib and Parasail compared to SHD. This is expected as SHD produces a high false accept rate (as we show earlier in Section \ref{filtering_accuracy}).
(4) The addition of SHD as a pre-alignment step reduces the execution time of Edlib and Parasail for some of the edit distance thresholds by up to 17.2$\times$ (using \texttt{100bp\_2}, $E=0\%$) and 34.9$\times$ (using \texttt{250bp\_2}, $E$= 3\%), respectively. 
However, for most of the edit distance thresholds, we observe that Edlib and Parasail are faster alone than with SHD combined as a pre-alignment filtering step. This is expected as SHD becomes ineffective in filtering for $E$> 8\% and $E$> 3\% for $m$= 100 bp and $m$= 250 bp, respectively, (as we show earlier in Section \ref{filtering_accuracy}).

We conclude that SneakySnake is the best-performing CPU-based pre-alignment filter in terms of both speed and accuracy. Integrating SneakySnake with sequence alignment algorithms is always beneficial for short sequences and reduces the end-to-end execution time by up to an order of magnitude without the need for hardware accelerators. 
We also conclude that SneakySnake's performance {scales well} over a wide range of edit distance thresholds, number of CPU threads, and sequence lengths.

\begin{figure}
\centerline{\includegraphics[width=1\columnwidth]{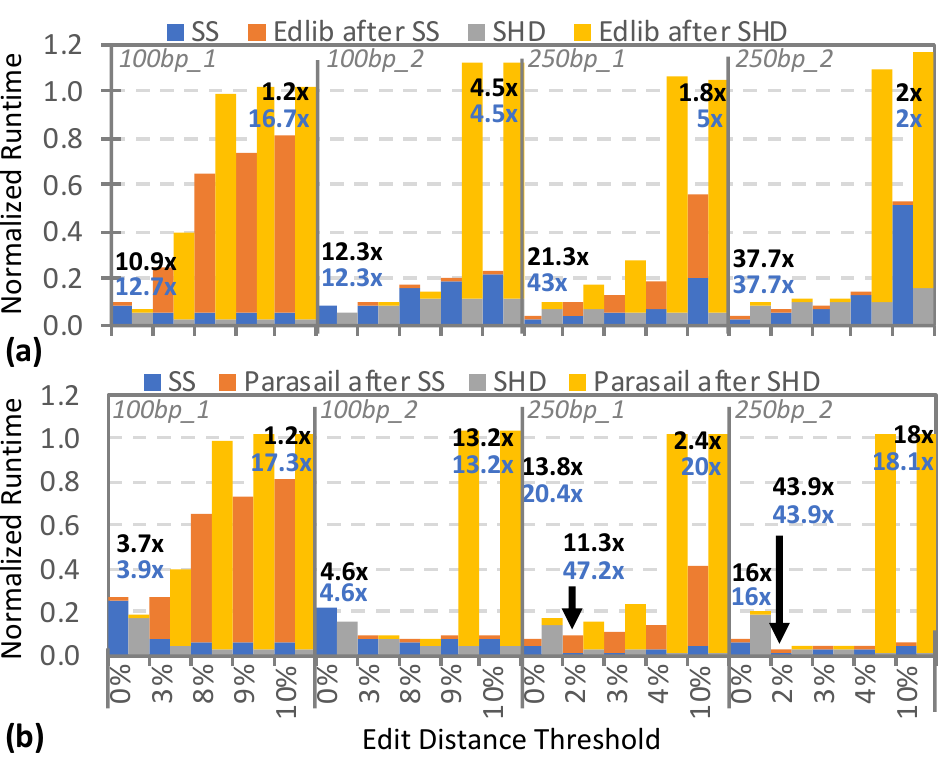}}
\caption{\alser{Normalized end-to-end execution time of SneakySnake and SHD{, each} combined with (a) Edlib and (b) Parasail. 
{The execution time values in (a) and (b) are normalized to that of Edlib and Parasail, respectively, {without} pre-alignment filtering.}
We use four datasets over a wide range of edit distance thresholds ($E$= 0\%-10\% of the sequence length) for sequence lengths ($m$) of 100 bp {(\texttt{100bp\_1} and \texttt{100bp\_2})} and 250 bp {(\texttt{250bp\_1} and \texttt{250bp\_2})}.
We present two speedup {values} for $E$= 0\% and $E$= 10\% of {each dataset} and some other $E$ values highlighted by arrows. The {top} speedup {value} {(in black)} represents the end-to-end speedup that is gained from combining the pre-alignment filtering step with the alignment step. It is calculated as $A/(B+C)$, where $A$ is the execution time of the sequence aligner before adding SneakySnake {(not plotted in graphs)}, $B$ is the execution time of SneakySnake, and $C$ is the execution time of the sequence aligner after adding SneakySnake. The {bottom} speedup {value} {(in blue)} is calculated as $A/B$.
}}
\label{fig:CPUSneakySnake}
\end{figure}

\subsection{{Effect} of Snake-on-Chip and Snake-on-GPU on Sequence Alignment}

We analyze the benefits of integrating Snake-on-Chip and Snake-on-GPU with the state-of-the-art sequence aligners, designed for different computing platforms in Fig. \ref{fig:Snake-on-Chip}.
{We} compare the effect of combining Snake-on-Chip and Snake-on-GPU with {an} existing sequence aligner {to} that of two state-of-the-art FPGA-based pre-alignment filters, Shouji and GateKeeper. 
We also select four state-of-the-art sequence aligners that are implemented for CPU (Edlib and Parasail), GPU (GSWABE), and FPGA (FPGASW). 
We use \texttt{100bp\_1} and \texttt{100bp\_2} in this {evaluation, as GSWABE, Shouji, and GateKeeper work for only short sequences}. GSWABE and FPGASW are not open-source and not available to us. Therefore, we scale {their} reported number of computed entries of the DP matrix {per} second (i.e., GCUPS) as follows: (number of sequence pairs in \texttt{100bp\_1} or \texttt{100bp\_2})/(GCUPS/$100^2$). 
We design the hardware architecture of Snake-on-Chip for a sub-maze's width of 8 VRTs ($t$=8) and 3 module instances ($y$=3) per each sub-maze. 
We select this design choice as it allows for low FPGA resource utilization while maintaining {a} low false accept rate{, based on our analysis of} different $y$ and $t$ values on the false accept rate of Snake-on-Chip {(these results} are reported in the {Supplementary Excel File} and on the SneakySnake GitHub page). 

Based on Fig. \ref{fig:Snake-on-Chip}, we make two key observations. (1) The execution time of Edlib and Parasail reduces by up to 321$\times$ (using \texttt{100bp\_2} and $E$ = 5\%) and 536$\times$ (using \texttt{100bp\_2} and $E$ = 5\%), respectively, after the addition of Snake-on-Chip as a pre-alignment filtering step and by up to 413$\times$ (using \texttt{100bp\_2} and $E$ = 5\%) and 689$\times$ (using \texttt{100bp\_2} and $E$ = 5\%), respectively, after the addition of Snake-on-GPU as a pre-alignment filtering step.
That is 40$\times$ (321/8) to 51$\times$ (689/13.39) more speedup {than} that provided by adding SneakySnake as a pre-alignment filter, using \texttt{100bp\_2} and $E$ = 5\%. 
It is also up to 2$\times$ more speedup compared to that provided by adding Shouji and GateKeeper as a pre-alignment filter, using \texttt{100bp\_1} and E=5\% for Snake-on-Chip and using \texttt{100bp\_2} and E=5\% for Snake-on-GPU. 
(2) 
Snake-on-GPU provides up to 27.7$\times$ (using \texttt{100bp\_2} and $E$ = 5\%) and 5.1$\times$ (using \texttt{100bp\_2} and $E$ = 5\%) reduction in the end-to-end execution time of GSWABE and FPGASW, respectively. 
This is up to 1.3$\times$ more speedup {than} that provided by Snake-on-Chip, using \texttt{100bp\_2}. That is also up to 1.7$\times$ more speedup {than} that provided by adding Shouji and GateKeeper as a pre-alignment filter. 
{The speedup provided by Snake-on-GPU and Snake-on-Chip to GSWABE and FPGASW is less than that observed in Edlib and Parasail. This is due to the low execution time of hardware accelerated aligners.}

We conclude that both Snake-on-Chip and Snake-on-GPU provide the highest speedup (up to two orders of magnitude) {when combined with} the state-of-the-art CPU, FPGA, and GPU based sequence aligners over edit distance thresholds of 0\%-5\% of the sequence length.

\begin{figure}
\centerline{\includegraphics[width=1\linewidth]{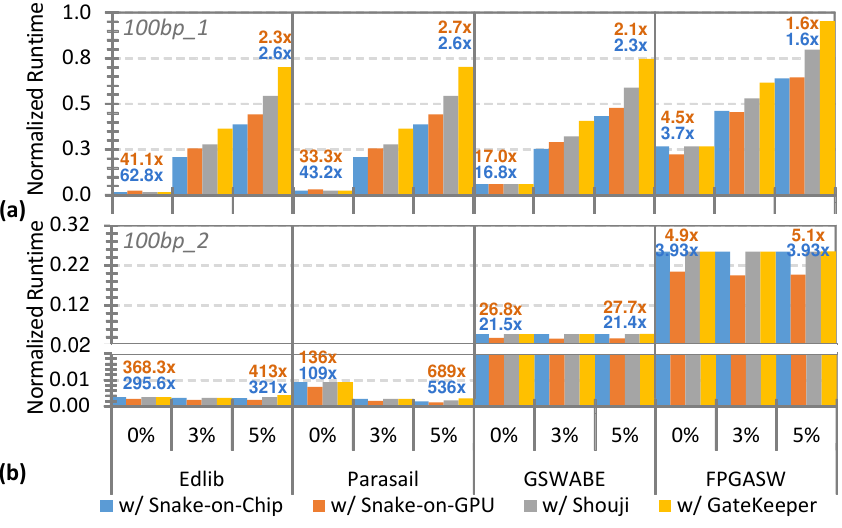}}
\caption{\alser{Normalized end-to-end execution time of a pre-alignment filter (Snake-on-Chip, Snake-on-GPU, Shouji, and GateKeeper) combined with a sequence aligner (Edlib, Parasail, GSWABE, and FPGASW). {Each execution time value is normalized to that of the corresponding sequence aligner {without} pre-alignment filtering.}
We use two datasets, (a) \texttt{100bp\_1} and (b) \texttt{100bp\_2}, over a wide range of edit distance thresholds (0\%-10\% of the sequence length, 100 bp). We present two end-to-end speedup {values} for edit distance thresholds of 0\% and 5\%. The {top} speedup {value} {(in orange)} is the speedup gained from integrating Snake-on-GPU with the corresponding sequence aligner. The {bottom} speedup {value} {(in blue)} represents the speedup gained from integrating Snake-on-Chip with the corresponding sequence aligner.}}
\label{fig:Snake-on-Chip}
\end{figure}

\subsection{{Effect} of SneakySnake on Long Sequence Alignment}
{We} examine the benefits of integrating SneakySnake with Parasail~\citep{daily2016parasail} and KSW2~\citep{suzuki2018introducing, li2018minimap2} {for long sequence alignment} (\texttt{100Kbp}). 
We run Parasail as nw\_banded. We run KSW2 as extz2\_sse, a global alignment implementation that is parallelized using the Intel SSE instructions{. KSW2} uses heuristics~\citep{suzuki2018introducing} to improve the alignment time. We run SneakySnake with Parasail using 40 CPU threads. We run SneakySnake with KSW2 using a single CPU thread (as KSW2 does not support multithreading). We use a wide range of edit distance thresholds, up to 20\% of the sequence length.

Based on Table \ref{table:1}, we make two key observations. (1) SneakySnake {accelerates} Parasail and KSW2 by 50.9-979$\times$ and 3.8-91.7$\times$, respectively, even at high edit distance thresholds (up to $E$=5010 (5\%), which results in building and examining a chip maze of 10,021 rows for each sequence pair).
(2) As the number of similar sequence pairs increases, the {performance benefit} of integrating SneakySnake with Parasail and KSW2 in reducing the end-to-end execution time {reduces}. 
{When Parasail and KSW2 examine 94\% and 73\% of the input sequence pairs (SneakySnake filters out the rest of the sequence pairs), respectively, SneakySnake provides slight or no performance benefit to the end-to-end execution time of the sequence aligner alone}. 
This is expected, as each sequence pair that passes SneakySnake is examined unnecessarily twice (i.e., once by SneakySnake and once by sequence aligner). We provide more details on this evaluation for both \texttt{10Kbp} and \texttt{100Kbp} in Supplementary Materials, Section 10.3.
We observe that SneakySnake accelerates Parasail and KSW2 by 276.9$\times$ and 31.7$\times$ on average, respectively, when {sequence alignment examines at most 73\% of the input sequence pairs}.

We conclude that {when SneakySnake filters out more than 27\% of the input sequence pairs, integrating SneakySnake with long sequence aligners is always beneficial and sometimes reduces the end-to-end execution time by one to two orders of magnitude (depending on the edit distance threshold and how fast the sequence aligner examines the input sequence pairs compared to SneakySnake) without the need for hardware accelerators}.

\begin{table}[H]
\caption{The end-to-end execution time (in seconds) of SneakySnake integrated with Parasail (40 CPU threads) and KSW2 (single threaded) using long reads (\texttt{100Kbp}).}
\label{table:1}
\centering
\begin{tabularx}{\linewidth}{c|rrrrr}
\hline
\textit{\textbf{E}} & \multicolumn{1}{l}{\textbf{Parasail}} & \multicolumn{1}{l}{\textbf{SS+Parasail}} & \multicolumn{1}{l}{\textbf{KSW2}} & \multicolumn{1}{l}{\textbf{SS+KSW2}} &  
\begin{minipage}[c] [.7cm] [c] {0.15\columnwidth}%
\centering
\textbf{SS Accept Rate} %
\end{minipage}
\\ \hline
0.01\% & 84.0 & \textbf{0.23} & 1380.2 & \textbf{15.1} & 0\% \\
0.3\% & 2,756.3 & \textbf{2.8} & 8,215.5 & \textbf{135.4} & 0\% \\
5.0\% & 37,492.3 & \textbf{736.5} & 100,178.3 & \textbf{26,261.4} & 0\% \\
10.7\% & 81,881.6 & \textbf{49,322.1} & 204,135.3 & \textbf{184,312.5} & 57\% \\
10.8\% & 82,646.1 & \textbf{63,756.0} & 206,041.4 & 225,815.2 & 73\% \\
11.0\% & 84,098.7 & \textbf{83,437.5} & 209,662.8 & 287,206.8 & 94\% \\
12.0\% & 91,744.1 & 95,533.6 & 228,723.1 & 325,966.0 & 100\% \\
20.0\% & 152,906.8 & 157,982.0 & 381,205.1 & 544,282.1 & 100\% \\ \hline
\end{tabularx}
\end{table}


\subsection{{Effect} of SneakySnake on Read Mapping}
\alser{After confirming the benefits of the different implementations of the SneakySnake algorithm, {we} evaluate the overall benefits of integrating SneakySnake with minimap2 (2.17-r974-dirty, 22 January 2020) \citep{li2018minimap2}. 
We select minimap2 for two main reasons. (1) It is a state-of-the-art read mapper that includes efficient methods (i.e., minimizers and seed chaining) for accelerating read mapping. (2) It utilizes a banded global sequence alignment algorithm (KSW2, implemented as \emph{extz2\_sse}) that is parallelized and accelerated using both the Intel SSE instructions and heuristics \citep{suzuki2018introducing} to improve the alignment time.
We map all reads from ERR240727\_1 (100 bp) to GRCh37 with edit distance thresholds of 0\% and 5\% of the sequence length. 
{We run minimap2 using --sr mode (short read mapping) and the default parameter values.}
We replace the seed chaining 
of minimap2 with SneakySnake.
In these experiments, we ensure that we maintain the \emph{same} reported mappings for both tools. 
We make two observations. 
(1) SneakySnake and the minimap2's aligner (KSW2) together are at least 6.83$\times$ (from 246 seconds to 36 seconds) and 2.51$\times$ (from 338 seconds to 134.67 seconds) faster than the minimap2's seed chaining and the minimap2's aligner together for edit distance thresholds of 0\% and 5\%, respectively. 
(2) The mapping time of minimap2 reduces by a factor of up to 2.01$\times$ (from 418 seconds to 208 seconds) and 1.66$\times$ (from 510 seconds to 306.67 seconds) after integrating SneakySnake with minimap2 for edit distance thresholds of 0\% and 5\%, respectively.}

{We conclude that SneakySnake is {very} beneficial even for minimap2, a state-of-the-art read mapper, which uses minimizers, seed chaining, and SIMD-accelerated banded alignment.
This promising result motivates us to explore in detail accelerating minimap2 using Snake-on-GPU and Snake-on-Chip in our future research.}

\section{Discussion and Future Work} \label{sec:conclusion}
We {demonstrate {that} we can} convert {the} approximate string matching problem into an instance of the single net routing problem. 
We {show how to do so and} propose a new algorithm that solves the single net routing problem and acts as a new pre-alignment filtering algorithm, {called} SneakySnake. 
SneakySnake offers the ability to make the best use of existing aligners without sacrificing any of their capabilities (e.g., configurable scoring functions and backtracking), as it does not modify or replace the alignment step. 
SneakySnake improves the accuracy of pre-alignment filtering by up to four orders of magnitude compared to {three} state-of-the-art pre-alignment filters, Shouji, GateKeeper, and SHD. The addition of SneakySnake as a pre-alignment filtering step {significantly} reduces the execution time of state-of-the-art CPU-based sequence aligners by up to an order {and two orders of magnitude using short and long sequences, respectively}. 
{We introduce Snake-on-Chip and Snake-on-GPU, efficient and scalable FPGA and GPU based hardware accelerators of SneakySnake, respectively}. Snake-on-Chip and Snake-on-GPU achieve up to {one order and} two orders of magnitude speedup {over} state-of-the-art {CPU- and hardware-based sequence aligners, respectively}.

One direction to further improve the performance of Snake-on-Chip is to discover the possibility of performing the SneakySnake calculations {near where huge amounts} of genomic data resides. Conventional computing requires the movement of genomic sequence pairs from the memory to the CPU processing cores (or to the {GPU or FPGA chips}), using slow and energy-hungry buses, such that cores can apply sequence alignment algorithm on the sequence pairs. 
Performing SneakySnake inside modern memory devices {via processing in memory}~\citep{mutlu2019processing, ghose2019processing} can alleviate this high communication cost by enabling simple arithmetic/logic operations very close to where the data resides, with high bandwidth, low latency{, and low energy}. However, this requires re-designing the hardware architecture of Snake-on-Chip to leverage the supported operations in such modern memory devices. 

\vspace{-6 pt}
\section*{Funding}
This work is supported by gifts from Intel [to O.M.]; VMware [to O.M.]; {a Semiconductor Research Corporation grant [to O.M.];} and an EMBO Installation Grant [IG-2521 to C.A.].\\
\vspace{-20 pt}

\bibliographystyle{natbib}

\bibliography{document}

\begin{thebibliography}{}

\bibitem[Alser {\em et~al.}(2017a)Alser, Hassan, Xin, Ergin, Mutlu, and
  Alkan]{alser2017gatekeeper}
Alser, M., Hassan, H., Xin, H., Ergin, O., Mutlu, O., and Alkan, C. (2017a).
\newblock {GateKeeper: a new hardware architecture for accelerating
  pre-alignment in DNA short read mapping}.
\newblock {\em Bioinformatics\/}, {\bf 33}(21), 3355--3363.

\bibitem[Alser {\em et~al.}(2017b)Alser, Mutlu, and Alkan]{alser2017magnet1}
Alser, M., Mutlu, O., and Alkan, C. (2017b).
\newblock {MAGNET: Understanding and improving the accuracy of genome
  pre-alignment filtering}.
\newblock {\em Transactions on Internet Research\/}, {\bf 13}(2), 33--42.

\bibitem[Alser {\em et~al.}(2019)Alser, Hassan, Kumar, Mutlu, and
  Alkan]{alser2019shouji}
Alser, M., Hassan, H., Kumar, A., Mutlu, O., and Alkan, C. (2019).
\newblock Shouji: a fast and efficient pre-alignment filter for sequence
  alignment.
\newblock {\em Bioinformatics\/}, {\bf 35}(21), 4255--4263.

\bibitem[Alser {\em et~al.}(2020a)Alser, Bing{\"o}l, Cali, Kim, Ghose, Alkan,
  and Mutlu]{alser2020accelerating}
Alser, M., Bing{\"o}l, Z., Cali, D.~S., Kim, J., Ghose, S., Alkan, C., and
  Mutlu, O. (2020a).
\newblock {Accelerating Genome Analysis: A Primer on an Ongoing Journey}.
\newblock {\em IEEE Micro\/}, {\bf 40}(5), 65--75.

\bibitem[Alser {\em et~al.}(2020b)Alser, Rotman, Taraszka, Shi, Baykal, Yang,
  Xue, Knyazev, Singer, Balliu, {\em et~al.}]{alser2020technology}
Alser, M., Rotman, J., Taraszka, K., Shi, H., Baykal, P.~I., Yang, H.~T., Xue,
  V., Knyazev, S., Singer, B.~D., Balliu, B., {\em et~al.} (2020b).
\newblock Technology dictates algorithms: Recent developments in read
  alignment.
\newblock {\em arXiv preprint arXiv:2003.00110\/}.

\bibitem[Chaisson and Tesler(2012)Chaisson and Tesler]{chaisson2012mapping}
Chaisson, M.~J. and Tesler, G. (2012).
\newblock {Mapping single molecule sequencing reads using basic local alignment
  with successive refinement (BLASR): application and theory}.
\newblock {\em BMC Bioinformatics\/}, {\bf 13}(1), 238.

\bibitem[Chakraborty {\em et~al.}(2018)Chakraborty, Das, Goldenberg, Koucky,
  and Saks]{chakraborty2018approximating}
Chakraborty, D., Das, D., Goldenberg, E., Koucky, M., and Saks, M. (2018).
\newblock {Approximating edit distance within constant factor in truly
  sub-quadratic time}.
\newblock In {\em IEEE Annual Symp. on Foundations of Computer Science
  (FOCS)\/}, pages 979--990.

\bibitem[Charikar {\em et~al.}(2018)Charikar, Geri, Kim, and
  Kuszmaul]{charikar2018estimating}
Charikar, M., Geri, O., Kim, M.~P., and Kuszmaul, W. (2018).
\newblock {On Estimating Edit Distance: Alignment, Dimension Reduction, and
  Embeddings}.
\newblock In {\em 45th International Colloquium on Automata, Languages, and
  Programming (ICALP)\/}.

\bibitem[Chen {\em et~al.}(2014)Chen, Wang, Li, and Zhou]{chen2014accelerating}
Chen, P., Wang, C., Li, X., and Zhou, X. (2014).
\newblock {Accelerating the next generation long read mapping with the
  FPGA-based system}.
\newblock {\em IEEE/ACM transactions on computational biology and
  bioinformatics\/}, {\bf 11}(5), 840--852.

\bibitem[Consortium {\em et~al.}(2015)Consortium {\em et~al.}]{10002015global}
Consortium, . G.~P. {\em et~al.} (2015).
\newblock A global reference for human genetic variation.
\newblock {\em Nature\/}, {\bf 526}(7571), 68--74.

\bibitem[Daily(2016)Daily]{daily2016parasail}
Daily, J. (2016).
\newblock {Parasail: SIMD C library for global, semi-global, and local pairwise
  sequence alignments}.
\newblock {\em BMC bioinformatics\/}, {\bf 17}(1), 81.

\bibitem[Eddy(2004)Eddy]{eddy2004dynamic}
Eddy, S.~R. (2004).
\newblock {What is dynamic programming?}
\newblock {\em Nature biotechnology\/}, {\bf 22}(7), 909.

\bibitem[Fei {\em et~al.}(2018)Fei, Dan, Lina, Xin, and Chunlei]{fei2018fpgasw}
Fei, X., Dan, Z., Lina, L., Xin, M., and Chunlei, Z. (2018).
\newblock {FPGASW: Accelerating Large-Scale Smith--Waterman Sequence Alignment
  Application with Backtracking on FPGA Linear Systolic Array}.
\newblock {\em Interdisciplinary Sciences: Computational Life Sciences\/}, {\bf
  10}(1), 176--188.

\bibitem[Firtina {\em et~al.}(2020)Firtina, Kim, Alser, Senol~Cali, Cicek,
  Alkan, and Mutlu]{firtina2020apollo}
Firtina, C., Kim, J.~S., Alser, M., Senol~Cali, D., Cicek, A.~E., Alkan, C.,
  and Mutlu, O. (2020).
\newblock {Apollo: a sequencing-technology-independent, scalable and accurate
  assembly polishing algorithm}.
\newblock {\em Bioinformatics\/}, {\bf 36}(12), 3669--3679.

\bibitem[Ghose {\em et~al.}(2019)Ghose, Boroumand, Kim, G{\'o}mez-Luna, and
  Mutlu]{ghose2019processing}
Ghose, S., Boroumand, A., Kim, J.~S., G{\'o}mez-Luna, J., and Mutlu, O. (2019).
\newblock {Processing-in-memory: A workload-driven perspective}.
\newblock {\em IBM Journal of Research and Development\/}, {\bf 63}(6), 3--1.

\bibitem[Kim {\em et~al.}(2018)Kim, Cali, Xin, Lee, Ghose, Alser, Hassan,
  Ergin, Alkan, and Mutlu]{kim2018grim}
Kim, J.~S., Cali, D.~S., Xin, H., Lee, D., Ghose, S., Alser, M., Hassan, H.,
  Ergin, O., Alkan, C., and Mutlu, O. (2018).
\newblock {GRIM-Filter: Fast seed location filtering in DNA read mapping using
  processing-in-memory technologies}.
\newblock {\em BMC Genomics\/}, {\bf 19}(2), 89.

\bibitem[Lee {\em et~al.}(1976)Lee, Bose, and Hwang]{lee1976use}
Lee, J., Bose, N., and Hwang, F. (1976).
\newblock {Use of Steiner's problem in suboptimal routing in rectilinear
  metric}.
\newblock {\em IEEE Transactions on Circuits and Systems\/}, {\bf 23}(7),
  470--476.

\bibitem[Levenshtein(1966)Levenshtein]{levenshtein1966binary}
Levenshtein, V.~I. (1966).
\newblock {Binary codes capable of correcting deletions, insertions, and
  reversals}.
\newblock In {\em Soviet Physics-Doklady\/}, volume~10, pages 707--710.

\bibitem[Li(2018)Li]{li2018minimap2}
Li, H. (2018).
\newblock {Minimap2: pairwise alignment for nucleotide sequences}.
\newblock {\em Bioinformatics\/}, {\bf 34}(18), 3094--3100.

\bibitem[Liu and Schmidt(2015)Liu and Schmidt]{liu2015gswabe}
Liu, Y. and Schmidt, B. (2015).
\newblock {GSWABE: faster GPU-accelerated sequence alignment with optimal
  alignment retrieval for short DNA sequences}.
\newblock {\em Concurrency and Computation: Practice and Experience\/}, {\bf
  27}(4), 958--972.

\bibitem[Mutlu {\em et~al.}(2019)Mutlu, Ghose, G{\'o}mez-Luna, and
  Ausavarungnirun]{mutlu2019processing}
Mutlu, O., Ghose, S., G{\'o}mez-Luna, J., and Ausavarungnirun, R. (2019).
\newblock {Processing data where it makes sense: Enabling in-memory
  computation}.
\newblock {\em Microprocessors and Microsystems\/}, {\bf 67}, 28--41.

\bibitem[Myers(1999)Myers]{myers1999fast}
Myers, G. (1999).
\newblock {A fast bit-vector algorithm for approximate string matching based on
  dynamic programming}.
\newblock {\em Journal of the ACM (JACM)\/}, {\bf 46}(3), 395--415.

\bibitem[Navarro(2001)Navarro]{navarro2001guided}
Navarro, G. (2001).
\newblock {A guided tour to approximate string matching}.
\newblock {\em ACM computing surveys (CSUR)\/}, {\bf 33}(1), 31--88.

\bibitem[Needleman and Wunsch(1970)Needleman and Wunsch]{needleman1970general}
Needleman, S.~B. and Wunsch, C.~D. (1970).
\newblock {A general method applicable to the search for similarities in the
  amino acid sequence of two proteins}.
\newblock {\em Journal of molecular biology\/}, {\bf 48}(3), 443--453.

\bibitem[NVIDIA(2019a)NVIDIA]{cuda-guide}
NVIDIA (2019a).
\newblock {{CUDA C} Programming Guide}.

\bibitem[NVIDIA(2019b)NVIDIA]{guide2019Nvidia2080}
NVIDIA (2019b).
\newblock {NVIDIA GeForce RTX 2080 Ti User Guide}.

\bibitem[Senol~Cali {\em et~al.}(2019)Senol~Cali, Kim, Ghose, Alkan, and
  Mutlu]{senol2019nanopore}
Senol~Cali, D., Kim, J.~S., Ghose, S., Alkan, C., and Mutlu, O. (2019).
\newblock {Nanopore sequencing technology and tools for genome assembly:
  computational analysis of the current state, bottlenecks and future
  directions}.
\newblock {\em Briefings in bioinformatics\/}, {\bf 20}(4), 1542--1559.

\bibitem[Senol~Cali {\em et~al.}(2020)Senol~Cali, Kalsi, Bing{\"o}l, Firtina,
  Subramanian, Kim, Ausavarungnirun, Alser, Luna, Boroumand, Nori, Scibisz,
  Subramoney, Alkan, Ghose, and Mutlu]{senolcalimicro2020}
Senol~Cali, D., Kalsi, G.~S., Bing{\"o}l, Z., Firtina, C., Subramanian, L.,
  Kim, J.~S., Ausavarungnirun, R., Alser, M., Luna, J.~G., Boroumand, A., Nori,
  A., Scibisz, A., Subramoney, S., Alkan, C., Ghose, S., and Mutlu, O. (2020).
\newblock {{GenASM: A High Performance, Low-Power Approximate String Matching
  Acceleration Framework for Genome Sequence Analysis}}.
\newblock In {\em MICRO\/}.

\bibitem[Seshadri {\em et~al.}(2017)Seshadri, Lee, Mullins, Hassan, Boroumand,
  Kim, Kozuch, Mutlu, Gibbons, and Mowry]{seshadri2017ambit}
Seshadri, V., Lee, D., Mullins, T., Hassan, H., Boroumand, A., Kim, J., Kozuch,
  M.~A., Mutlu, O., Gibbons, P.~B., and Mowry, T.~C. (2017).
\newblock {Ambit: In-memory accelerator for bulk bitwise operations using
  commodity DRAM technology}.
\newblock In {\em MICRO\/}.

\bibitem[{\v{S}}o{\v{s}}i{\'c} and {\v{S}}iki{\'c}(2017){\v{S}}o{\v{s}}i{\'c}
  and {\v{S}}iki{\'c}]{vsovsic2017edlib}
{\v{S}}o{\v{s}}i{\'c}, M. and {\v{S}}iki{\'c}, M. (2017).
\newblock {Edlib: a C/C++ library for fast, exact sequence alignment using edit
  distance}.
\newblock {\em Bioinformatics\/}, {\bf 33}(9), 1394--1395.

\bibitem[Suzuki and Kasahara(2018)Suzuki and Kasahara]{suzuki2018introducing}
Suzuki, H. and Kasahara, M. (2018).
\newblock Introducing difference recurrence relations for faster semi-global
  alignment of long sequences.
\newblock {\em BMC bioinformatics\/}, {\bf 19}(1), 33--47.

\bibitem[Wang {\em et~al.}(2011)Wang, Yan, Wang, Si, and
  Zhang]{wang2011comparison}
Wang, C., Yan, R.-X., Wang, X.-F., Si, J.-N., and Zhang, Z. (2011).
\newblock {Comparison of linear gap penalties and profile-based variable gap
  penalties in profile--profile alignments}.
\newblock {\em Computational biology and chemistry\/}, {\bf 35}(5), 308--318.

\bibitem[Xilinx(2013)Xilinx]{guide2013virtex}
Xilinx (2013).
\newblock {Virtex-7 XT VC709 Connectivity Kit}.

\bibitem[Xin {\em et~al.}(2013)Xin, Lee, Hormozdiari, Yedkar, Mutlu, and
  Alkan]{xin2013accelerating}
Xin, H., Lee, D., Hormozdiari, F., Yedkar, S., Mutlu, O., and Alkan, C. (2013).
\newblock {Accelerating read mapping with FastHASH}.
\newblock In {\em BMC genomics\/}, volume~14, page S13.

\bibitem[Xin {\em et~al.}(2015)Xin, Greth, Emmons, Pekhimenko, Kingsford,
  Alkan, and Mutlu]{xin2015shifted}
Xin, H., Greth, J., Emmons, J., Pekhimenko, G., Kingsford, C., Alkan, C., and
  Mutlu, O. (2015).
\newblock {Shifted Hamming distance: a fast and accurate SIMD-friendly filter
  to accelerate alignment verification in read mapping}.
\newblock {\em Bioinformatics\/}, {\bf 31}(10), 1553--1560.

\end{thebibliography}

\includepdf[pages=-]{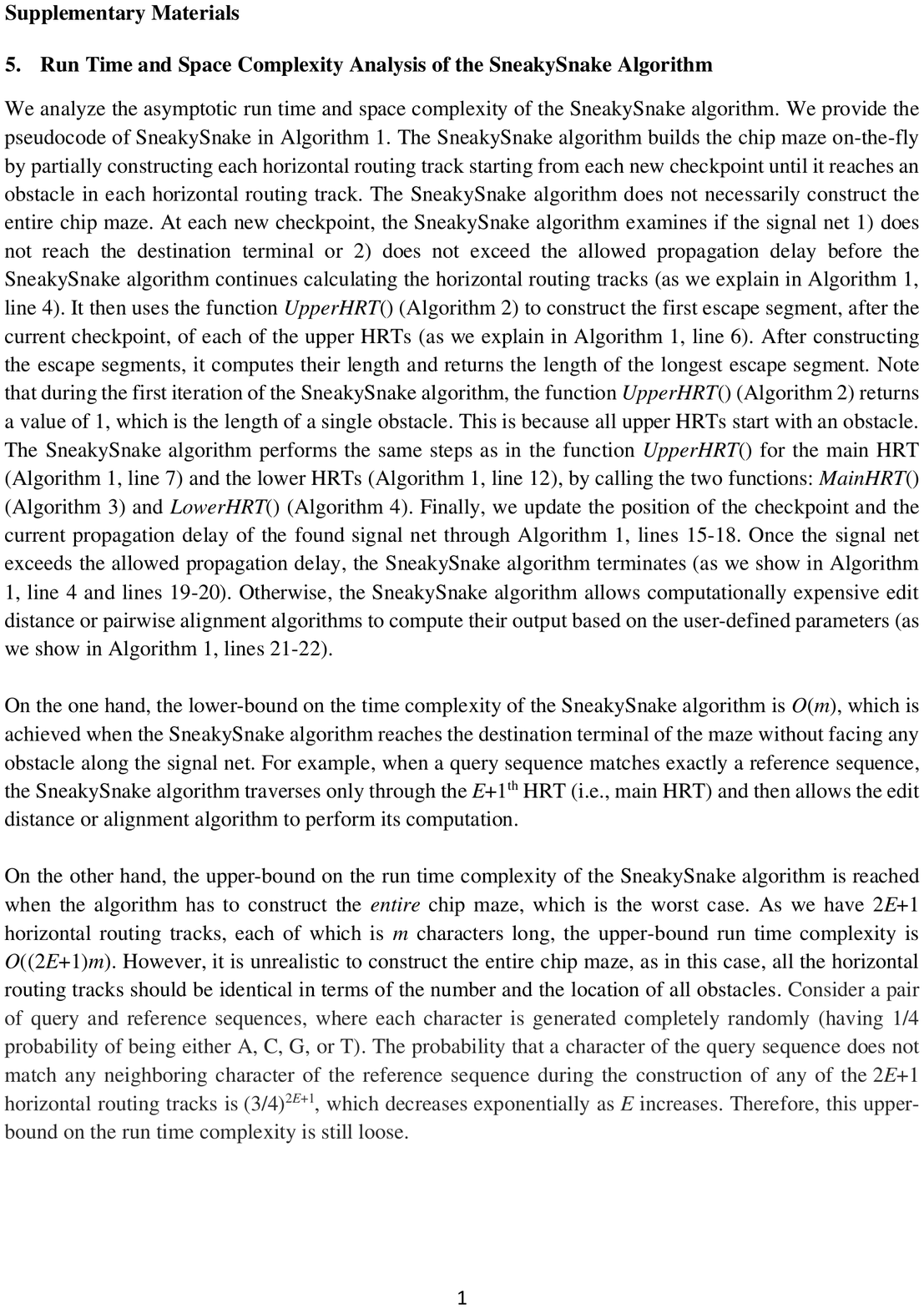}

\end{document}


\maketitle

\section{Constructing a profile hidden Markov model} \label{suppsec:phmm}

A pHMM-graph includes states and directed transitions from a state to another. There are two type of probabilities that the graph contains: (1) emission and (2) transition probabilities. First, each state has emission probabilities for emitting the certain characters where each is associated with a probability in range $[0,1]$. Second, each transition is associated with a probability in range $[0,1]$. Now, we will explain the structure of the graph in detail. For an assembly contig $C$, let us define the pHMM-graph that represents the contig $C$ as $G(V,E)$. Let us also define the length of the contig $C$ as $n = |C|$. A basepair $C[t]$  has one of the letters in the alphabet set $\Sigma=\{A,C,G,T\}$. Thus, a state emits one of the characters in $\Sigma$ with a certain probability. For a state $i$, We denote the emission probability of a basepair $c \in \Sigma$ as $e_{i}(c) \in [0,1]$ where $\sum\limits_{c \in \Sigma} e_{i}(c) = 1$. We denote the transition probability from a state, i, to another state, j, as $\alpha_{ij} \in [0,1]$. For the set of the states that the state $i$ has an outgoing transition to, $V_{i}$, we have $\sum\limits_{j \in V_{i}} \alpha_{ij} = 1$. Now let us define in four steps how Apollo constructs the states and the transitions of the graph $G(V,E)$:

First, Apollo constructs a start state, $v_{start} \in V$, and an end state $v_{end} \in V$. Second, for each basepair $C[t]$ where $1 \leq t \leq n$, Apollo constructs a match state as follows (Figure~\ref{fig:match}):
\begin{itemize}
    \item A \textit{match state} that we denote as $M_{t}$ for the basepair $C[t]$ where $M = C[t]$ s.t. $C[t] \in \Sigma$ and $M_{t} \in V$ (i.e., if the $t_{th}$ basepair of the contig $C$ is $G$, then the corresponding match state is $G_t$). For the following steps, let us assume $i = M_{t}$
    \item A \textit{match emission} with the probability $\beta$, for the basepair $C[t]$ s.t. $e_i(C[t]) = \beta$. $\beta$ is a parameter to Apollo. 
    \item A \textit{substitution emission} with the probability $\delta$, for each basepair $c \in \Sigma$ and $c \neq C[t]$ s.t. $e_i(c) = \delta$ (Note that $\beta + 3\delta = 1$). $\delta$ is a parameter to Apollo.
    \item A \textit{match transition} with the probability $\alpha_{M}$, from the match state $M_t = i$ to the next match state $M_{t+1} = j$ s.t. $\alpha_{ij} = \alpha_{M}$. $\alpha_{M}$ is a parameter to Apollo.
\end{itemize}

\begin{figure}
\centerline{\includegraphics[scale=0.4]{figures/MatchAndSubs_small.png}}
\caption{Two match states. Here, the contig includes the basepairs $G$ and $A$ at the locations $t$ and $t+1$, respectively. The corresponding match states are labeled with the basepairs that they correspond to (i.e., the match state $G_{t}$ represents the basepair $G$ at the location $t$). Each match state has a match transition to the next match state with the initial probability $\alpha_{M}$. A match state has a match emission probability, $\beta$, for the basepair it is labeled with. The remaining three basepairs have equal substitution emission probability $\delta$. The figure is taken from Hercules \cite{Firtina2018}.}
\label{fig:match}
\end{figure}

Third, for each basepair $C[t]$ where $1 \leq t \leq n$, Apollo constructs the insertion states as follows (Figure~\ref{fig:insertion}):
\begin{itemize}
    \item There are $l$ many \textit{insertion states}, $I_{t}^{1}$, $I_{t}^{2}$, \dots, $I_{t}^{l}$, where $I_{t}^{i} \in V$, $1 \leq i \leq l$ and $l$ is a parameter to Apollo
    \item The match state, $M_{t} = i$, has an \textit{insertion transition} to $I_{t}^{1} = j$, with the probability $\alpha_{I}$ s.t. $\alpha_{ij} = \alpha_{I}$
    \item For each $i$ where $1 \leq i < l$, the insertion state $I_{t}^{i} = k$ has an insertion transition to the next insertion state $I_{t}^{i+1} = j$ with the probability $\alpha_{I}$ s.t. $\alpha_{kj} = \alpha_{I}$
    \item For each $i$ where $1 \leq i < l$, the insertion state $I_{t}^{i} = k$ has a match transition to the match state of the next basepair $M_{t+1} = j$ with the probability $\alpha_{M}$ s.t. $\alpha_{kj} = \alpha_{M}$
    \item The last insertion state, $I_{t}^{l}$, has no further insertion transitions. Instead, it has a transition to the match state of the next basepair $M_{t+1} = j$ with the probability $\alpha_{M} + \alpha_{I}$ s.t. $\alpha_{kj} = \alpha_{M} + \alpha_{I}$
    \item For each $i$ where $1 \leq i \leq l$, each basepair $c \in \Sigma$ and $c \neq C[t+1]$ has an \textit{insertion emission} probability $1/3 \approx 0.33$ for the insertion state $I_{t}^{i} = k$ s.t. $e_k(c) = 0.33$ and $e_k(C[t+1]) = 0$. Note that $\sum\limits_{c \in \Sigma} e_{k}(c) = 1$. (i.e., if the basepair at the location $t+1$ is T, then $e_{k}(A) = 0.33$, $e_{k}(T) = 0$, $e_{k}(G) = 0.33$, and $e_{k}(C) = 0.33$).
\end{itemize}

\begin{figure}
\centerline{\includegraphics[scale=0.4]{figures/Insertions_small.png}}
\caption{$l$ many insertion states for the basepair at location $t$. Here, the contig includes the basepairs $C$ and $T$ at the locations $t$ and $t+1$, respectively. The corresponding match states are labeled with the basepairs that they correspond to. Each insertion state has an insertion transition to the next insertion state with the initial probability $\alpha_{I}$ and a match transition to the next match state at the location $t+1$ with the initial probability $\alpha_{M}$. However, the last insertion state, $I_{t}^{l}$, does not have a transition to the next insertion state as it is the last one. Instead, it has a match transition to the next match state $T_{t+1}$ with the probability $\alpha_{M} + \alpha_{I}$. The emission probability of the basepair $T$ is $0$ as it appears in the next position ($t+1$) of the contig. The figure is taken from Hercules \cite{Firtina2018}.}
\label{fig:insertion}
\end{figure}

Fourth step for finalizing the complete structure of the pHMM graph, for each \textit{state} $i \in V$, Apollo constructs the deletion transitions as follows (Figure~\ref{fig:deletion}):

\begin{itemize}
    \item Let us define $\alpha_{del} = 1 - (\alpha_{M} - \alpha_{I})$, which is the overall deletion transition probability.
    \item There are $k$ many deletion transitions from the state $i$, to the further \textit{match} states. $k$ is a parameter to Apollo.
    \item We assume that a transition deletes the basepairs if it skips the corresponding match states of the basepairs. We denote the transition probability of a deletion transition as $\alpha_{D}^{x}$ s.t. $1 \leq x \leq k$, if it deletes $x$ many basepairs in a row in one transition. Apollo calculates the deletion transition probability $\alpha_{D}^{x}$ using the normalized version of a polynomial distribution where $f \in [0, \infty)$ is a factor value for the equation:
    
    \begin{equation}
        \alpha_{D}^{x} = \dfrac{f^{k-x}\alpha_{del}}{\sum\limits_{j=0}^{k-1} f^{j}}\quad 1 \leq x     \leq k \tag{S1} \label{eq:deletion}
    \end{equation}
    
    \item If the $f$ value is set to $1$, then the each deletion transition is equally likely (i.e., $\alpha_{D}^{1} = \alpha_{D}^{10}$, if $k \geq 10$). As the $f$ value increases, the probability of deleting more basepairs in one transition decreases accordingly (i.e., $\alpha_{D}^{1} \gg \alpha_{D}^{10}$, if $k \geq 10$). $f$ is a parameter to Apollo.
\end{itemize}

\begin{figure}
\centerline{\includegraphics[scale=0.5]{figures/Deletions_small.png}}
\caption{Deletion transitions of the match and each insertion states at location $t$. For the match and insertion states at location $t$, we show only the deletion transitions (red). Note that a deletion transition from the position $t$ to the match state of the position $t+x+1$ removes $x$ many basepairs with the probability $\alpha_{D}^{x}$ as it skips $x$ many match states where $1 \leq x \leq k$. The figure is taken from Hercules \cite{Firtina2018}.}
\label{fig:deletion}
\end{figure}

We note that the start state $v_{start}$ also has a match transition to $M_{1}$ and deletion transitions as defined previously. There are al $l$ many insertion states, $I_{0}^{1}$, $I_{0}^{2}$, \dots, $I_{0}^{l}$, between the start state and the first match state $M_{1}$. The transitions of these insertion states are also identical to what we described before. We would also like to note that the end state $v_{end}$ has no outgoing transition. The prior states consider $v_{end}$ as a match state and connect to it accordingly. The start and end states have no emission probabilities.

\section{The Forward-Backward Algorithm}\label{suppsec:fwbw}

Apollo trains the pHMM of the contig per each read that aligns to the contig. It uses the alignment location and the sequence of the read in order to train the pHMM-graph. First, per each aligned read sequence $r$, Apollo extracts the sub-graph $G_s(V_s,E_s)$ that corresponds to the aligned region of the contig where we have $v_{start}$, $v_{end}$, match and insertion states, and the transitions as described in the Supplementary Section~\ref{suppsec:phmm}. Each transition from state $i \in V_{s}$ to state $j \in V_{s}$, $E_{ij} \in E_s$, is associated with a transition probability $\alpha_{ij}$. For every pair of states, $i \in V_s$ and $j \in V_s$, the transition probability $\alpha_{ij} = 0$ if $E_{ij} \not\in E_s$. Let us define the length of the aligned read, $r$, as $m = |r|$. Second, it calculates the forward and backward probabilities of each state based on the aligned read, $r$.

Let us assume that the forward probability of a state $j$ that observes $t^{th}$ basepair of the aligned read, $r[t]$, is $F_{t}(j)$. For the forward probability, observing the $t^{th}$ basepair at the state $j$ means that all the previous basepairs ($r[1] \dots r[t-1]$ and $1 < t \leq m$) have been observed by following a path starting from the start state to the state $j$ and $j$ observes the next basepair, $r[t]$. All possible transitions that lead to state $j$ to observe the basepair $r[t]$ contribute to the probability with (1) the forward probability of the origin state $i$ calculated with the $(t-1)^{th}$ basepair of $r$, $F_{t-1}(i)$, (2) multiplied by the probability of the transition from $i$ to $j$, $\alpha_{ij}$, (3) multiplied by the probability of emitting the basepair $r[t]$ at state $j$, $e_j(r[t])$.

Let us denote the start state $v_{start}$ with the index value of $0$ (i.e., $v_{start} = 0$). For each state $j \in V_{s}$, we calculate the forward probability, $F_{t}(j)$, as follows where $F_{1}(j)$ is the initialization step:

\begin{equation}
F_{1}(j) = \alpha_{0j}e_j(r[1]) \quad s.t. \quad j \in V_{s}, \quad E_{0j} \in E_s \tag{S2.1} \label{suppeq:initforward}
\end{equation}

\begin{equation}
F_{t}(j) = \sum_{i \in V_{s}} F_{t-1}(i)\alpha_{ij}e_j(r[t]) \quad j \in V_{s}, \quad 1 < t \leq m \tag{S2.2} \label{suppeq:forward}
\end{equation}

Let us assume that the backward probability of a state $i$ that observes $t^{th}$ basepair of the aligned read, $r[t]$, is $B_{t}(i)$. For the backward probability, observing the $t^{th}$ basepair at the state $i$ means that all the further basepairs ($r[t+1] \dots r[m]$ and $1 \leq t < m$) have been observed by following a path starting from the end state to the state $i$ (backwards) and $i$ observes the previous basepair, $r[t]$. All possible transitions that lead to state $i$ to observe the basepair $r[t]$ contribute to the probability with (1) the backward probability of the next state $j$ calculated with the $(t+1)^{th}$ basepair of $r$, $B_{t+1}(j)$, (2) multiplied by the probability of the transition from $i$ to $j$, $\alpha_{ij}$, (3) multiplied by the probability of emitting the basepair $r[t+1]$ at state $j$, $e_j(r[t+1])$.

Let us denote the end state $v_{end}$ with the index value of $m+1$ (i.e., $v_{end} = m+1$). For each state $j \in V_{s}$, we calculate the backward probability, $B_{t}(i)$, as follows where $B_{m}(i)$ is the initialization step:

\begin{equation}
B_{m}(i) = \alpha_{i(m+1)}\quad i \in V_{s}, \quad E_{i(m+1)} \in E_s\tag{S3.1} \label{suppeq:initbackward}
\end{equation}
\begin{equation}
B_{t}(i) = \sum_{j \in V_{s}} \alpha_{ij}e_j(r[t+1])B_{t+1}(j) \quad j \in V_{s}, ~1 \leq t < m \tag{S3.2} \label{suppeq:backward}
\end{equation}

After calculation of the forward and backward probabilities, Apollo calculates the posterior transition and the emission probabilities of the sub-graph, $G_s$, as shown in equations~\ref{suppeq:emissionupd} and~\ref{suppeq:transitionupd}, respectively. 

\begin{equation}
e^{*}_i(X) = \dfrac{\sum\limits_{t=1}^m F_{t}(i)B_t(i)(r[t] == X)}
             {\sum\limits_{t=1}^m F_{t}(i)B_t(i)} \quad \forall X \in \Sigma, \forall i \in V_{s} \tag{S4} \label{suppeq:emissionupd}
\end{equation}

\begin{equation}
\alpha^{*}_{ij} = \dfrac{\sum\limits_{t=1}^{m-1} \alpha_{ij}e_{j}(r[t+1])F_{t}(i)B_{t+1}(j)}
     {\sum\limits_{t=1}^{m-1}\sum\limits_{x \in V_{s}} \alpha_{ix}e_{x}(r[t+1])F_{t}(i)B_{t+1}(x)} \quad \forall E_{ij} \in E_{s}
\tag{S5} \label{suppeq:transitionupd} 
\end{equation}

\section{Joining Posterior Probabilities}\label{suppsec:join}
As we explain in the Supplementary Section~\ref{suppsec:fwbw}, for each read that aligns to the contig, Apollo extracts a sub-graph $G_s$ and uses the Forward-Backward algorithm to train the sub-graph. It is possible that there can be overlaps between two or many sub-graphs such that the sub-graphs can include the same states and the transitions. However, the updates on the overlapping states and the transitions are exclusive between the sub-graphs such that no two update in separate graphs affect each other while calculating the Forward or the Backward probabilities. Each sub-graph uses the initial probabilities to calculate the posterior probabilities. In order to handle training of the overlapping states and the transitions, Apollo takes the average of the posterior probabilities and reports the average probability as the final posterior probability for the entire pHMM-graph.

Let us assume that the set of sub-graphs $S$ includes the same state $i \in V$. For each $G_{s}$ in $S$, we obtain a $e^{*}_i(X)$, where $\forall X \in \Sigma$, which denotes the posterior emission probability as we explain in the Supplementary Section~\ref{suppsec:fwbw}. We denote $e^*_i(X)$ that belongs to $G_{s}$ as $e^{*,G_{s}}_i(X)$. Then, Apollo finds the final emission value $\hat{e}_i(X)$ as follows: 

\begin{equation}
\hat{e}_i(X) = \dfrac{\sum\limits_{G_{s} \in S} e^{*,G_{s}}_{i}(X)}{\mid S \mid} \quad \forall X \in \Sigma
\tag{S6}\label{suppeq:emissioncapupd}
\end{equation}

Similarly, let us assume that the set of sub-graphs $S$ includes the same transition edge $E_{ij} \in E$.  For each $G_{s}$ in $S$, we obtain an $\alpha^{*}_{ij}$ that denotes the posterior transition value. We define $\alpha^{*}_{ij}$ that belongs to $G_{s}$ as $\alpha^{*,G_{s}}_{ij}$. Apollo finds the final transition value $\hat{\alpha}_{ij}$ as follows: 

\begin{equation}
\hat{\alpha}_{ij} = \dfrac{\sum\limits_{G_{s} \in S} \alpha^{*,G_{s}}_{ij}}{\mid S \mid}
\tag{S7} \label{suppeq:transitioncapupd} 
\end{equation}

If a state in $V$ or an edge in $E$ is not covered by a read then Apollo retains the initial emission and transition probabilities and uses as posterior probabilities, respectively.

\section{Decoding with the Viterbi Algorithm}\label{suppsec:decoding}

Apollo uses the Viterbi algorithm to reveal the polished assembly by finding the most likely path starting from the start state, $v_{start}$ of the graph $G$ to the end state $v_{end}$. For each state $j$, the Viterbi algorithm calculates $v_{t}(j)$, which is the maximum marginal forward probability $j$ obtained from following a path starting from the start state when emitting the $t^{th}$ basepair of the polished contig. It also keeps a back pointer, $b_{t}(j)$, which keeps track of the predecessor state $i$ that yields the $v_{t}(j)$ value.

Let $X' \in \Sigma$ be the basepair that has the highest posterior emission probability for the state $j$ and $T$ be the length of the decoded sequence, which is initially unknown. The algorithm recursively calculates $v$ values for each position $t$ of a decoded sequence as described in the equations ~\ref{suppeq:vitinitial} and ~\ref{suppeq:vitrec}. The algorithm stops at iteration $T^{*}$ such that for the last $iter$ iterations, the maximum value we have observed for $v(end)$ cannot be improved and $iter$ is a parameter and set to 100 by default. $T$ is then set to $t^{*}$ such that $v_{t^{*}}(end)$ is the maximum among all iterations $1 \leq t \leq T^{*} $.

\begin{enumerate}
\item Initialization
\begin{equation}
v_{1}(j) = \hat{a}_{start-j}\hat{e}_{j}(X') \quad \forall j \in V \tag{S8.1} \label{suppeq:vitinitial} \\
\end{equation}

\begin{equation}
b_{1}(j) = start \tag{S8.2} 
\quad \forall j \in V \label{suppeq:backtraceinitial}
\end{equation}

\item Recursion
\begin{equation}
v_{t}(j) = \max_{i \in V} v_{t-1}(i)\hat{\alpha}_{ij}\hat{e}_{j}(X') \quad \forall j \in V, 1 < t \leq T \tag{S8.3} \label{suppeq:vitrec}
\end{equation}

\begin{equation}
b_{t}(j) = \argmax_{i \in V} v_{t-1}(i)\hat{\alpha}_{ij}\hat{e}_{j}(X') \quad \forall j \in V, 1 < t \leq T  \tag{S8.4} \label{suppeq:backtracerec}
\end{equation}

\item Termination
\begin{equation}
v_{T}(end) = \max_{i \in V} v_{T}(i)\hat{\alpha}_{i-end} \tag{S8.5} \label{suppeq:vitterm}
\end{equation}

\begin{equation}
b_{T}(end) = \argmax_{i \in V} v_{T}(i)\hat{\alpha}_{i-end} \tag{S8.6} \label{suppeq:backtraceterm}
\end{equation}
\end{enumerate}

\clearpage

\section{Performance of the Assembly Polishing Algorithms}

In Tables ~\ref{supptab:ecolik12} and ~\ref{supptab:chm1}, we compare the assembly polishing performance of Apollo with the competing algorithms.

\begin{table}[htb]
\begin{center}
\caption{Assembly polishing performance of the tools for E.Coli K-12 data set}
\label{supptab:ecolik12}
\resizebox{\textwidth}{!}{
\begin{tabular}{|l|l|l|l|l||rrr|rr|}
\hline Sequencing Tech. & Assembler & Aligner & Sequencing Tech. & Polishing & Aligned & Coverage & Avg. & Run Time & Memory\\
of the Assembly & & & of the Aligned Reads & Algorithm & Bases & (\%) & Identity (\%) & & (GB) \\\hline
ONT & Miniasm & - & - & - & 4,103,234 & 86.43 & 85.15 & 14m 45s & \\
ONT & Miniasm & Minimap2 & ONT & Apollo & 4,552,848 & 96.97 & 91.16 & 26h 34m 02s & 25.45 \\
ONT & Miniasm & Minimap2 & ONT & Nanopolish & 4,616,602 & 94.36 & 91.67 & 301h 42m 14s & 8.98 \\
ONT & Miniasm & Minimap2 & ONT & Racon & 4,911,410 & 99.22 & \textbf{97.66} & \textbf{15m 23s} & \textbf{3.65} \\\hhline{|=|=|=|=|=|===|==|}
ONT & Canu & - & - & - & 4,611,653 & 99.98 & 97.93 & 35h 36m 58s & \\
ONT & Canu & Minimap2 & ONT & Apollo & 4,613,425 & 99.99 & 97.99 & 68h 26m 55s & 31.39 \\
ONT & Canu & Minimap2 & ONT & Nanopolish & 4,689,609 & 99.99 & \textbf{99.24} & 19h 49m 10s & \textbf{3.20} \\
ONT & Canu & Minimap2 & ONT & Racon & 4,656,091 & 100.00 & 98.40 & \textbf{19m 07s} & 3.81 \\\hhline{|=|=|=|=|=|===|==|}
ONT (30X) & Canu & - & - & - & 3,771,861 & 100.00 & 97.11 & 27m 23s & \\
ONT (30X) & Canu & Minimap2 & ONT (30X) & Apollo & 3,771,288 & 99.99 & 97.14 & 2h 03m 09s & 1.37 \\
ONT (30X) & Canu & Minimap2 & ONT (30X) - Corrected & Apollo & 3,784,380 & 99.99 & 97.50 & 2h 48m 43s & 1.29 \\
ONT (30X) & Canu & Minimap2 & ONT (30X) & Nanopolish & 3,829,723 & 100.00 & \textbf{98.39} & 5h 25m 50s & 1.30 \\
ONT (30X) & Canu & Minimap2 & ONT (30X) & Racon & 3,827,938 & 100.00 & 98.30 & 45s & 0.42 \\
ONT (30X) & Canu & Minimap2 & ONT (30X) - Corrected & Racon & 3,811,164 & 100.00 & 98.09 & \textbf{20s} & \textbf{0.33}\\\hline
\end{tabular}}
\end{center}
{\footnotesize We generate the assembly using the reads (E.Coli K-12) sequenced from Oxford Nanopore Technologies (ONT) (319X coverage) as specified in \textit{Sequencing Tech. of the Assembly}. We subsample ONT reads into 30X coverage and generate the assembly using the sub-sampled reads that we show as "ONT (30X)". We use Canu and Miniasm assemblers as specified under \textit{Assembler}. Here, the reads specified under \textit{Sequencing Tech. of the Aligned Reads} are sequenced by the specified sequencing technology and are aligned to the assembly using the \textit{Aligner}. Canu-corrected long reads are labeled as "Corrected". We report the performance of the tools in terms of the number of bases aligned to the reference (\textit{Aligned Bases}), the percentage of the whole assembly that can align to the reference (\textit{Coverage}), and its \textit{Average Identity} (accuracy) as calculated by dnadiff.  We report the running time and the memory requirements of the assembly polishing tools. We report the performance of the assemblers in the rows where we do not specify a \textit{Polishing Algorithm}. We highlight the best result in each performance metric.}
\end{table}

\begin{table}[htb]
\begin{center}
\caption{Assembly polishing performance of the tools for CHM1 (Homo Sapiens) data set}
\label{supptab:chm1}
\resizebox{\textwidth}{!}{
\begin{tabular}{|l|l|l|l|l||rrr|rr|}
\hline Sequencing Tech. & Assembler & Aligner & Sequencing Tech. & Polishing & Aligned & Coverage & Avg. & Run Time & Memory\\
of the Assembly & & & of the Aligned Reads & Algorithm & Bases & (\%) & Identity (\%) & & (GB) \\\hline
PacBio & Miniasm & - & - & - & 1,800,151 & 50.30 & 79.92 & 8h 52m 43s & \\
PacBio & Miniasm & Minimap2 & PacBio & Apollo & 1,868,006 & 53.65 & \textbf{81.79} & 6h 58m 05s & \textbf{2.99} \\
PacBio & Miniasm & Minimap2 & PacBio & Racon & 1,882,827 & 57.29 & 78.50 & \textbf{16m 46s} & 15.42\\\hhline{|=|=|=|=|=|===|==|}
PacBio & Miniasm & pbalign & PacBio & Quiver & 1,918,464 & 54.13 & \textbf{82.99} & \textbf{3m 58s} & \textbf{0.50} \\\hhline{|=|=|=|=|=|===|==|}
PacBio & Canu & - & - & - & 1,891,967 & 89.14 & \textbf{90.31} & 64h 10m 02s & \\
PacBio & Canu & Minimap2 & PacBio & Apollo & 1,875,435 & 88.85 & 89.49 & 10h 45m 01s & 2.61  \\
PacBio & Canu & Minimap2 & PacBio - Corrected & Apollo & 1,880,259 & 89.04 & 89.77 & 11h 07m 35s & \textbf{2.34} \\
PacBio & Canu & Minimap2 & PacBio & Racon & 1,929,488 & 89.93 & 87.44 & 1h 15m 10s & 13.41 \\
PacBio & Canu & Minimap2 & PacBio - Corrected & Racon & 1,909,958 & 90.04 & 87.95 & \textbf{1h 12m 37s} & 11.83\\\hhline{|=|=|=|=|=|===|==|}
PacBio & Canu & pbalign & PacBio & Quiver & 1,900,762 & 89.48 & 89.25 & \textbf{1m 54s} & \textbf{0.53} \\\hline
\end{tabular}}
\end{center}
{\footnotesize We generate the assembly using the reads (CHM1) sequenced from PacBio (2.6X coverage) as specified in \textit{Sequencing Tech. of the Assembly}. We use Canu and Miniasm assemblers as specified in \textit{Assembler}. Here, the reads specified under \textit{Sequencing Tech. of the Aligned Reads} are sequenced by the specified sequencing technology and are aligned to the assembly using the \textit{Aligner}. Canu-corrected long reads are labeled as "Corrected". We report the performance of the tools in terms of the number of bases aligned to the reference (\textit{Aligned Bases}), the percentage of the whole assembly that can align to the reference (\textit{Coverage}), and its \textit{Average Identity} (accuracy) as calculated by BLASR. We report the running time and the memory requirements of the assembly polishing tools. We report the performance of the assemblers in the rows where we do not specify a \textit{Polishing Algorithm}. We highlight the best result in each performance metric.}
\end{table}

\clearpage

\section{Performance of the Aligners}

Here in Table ~\ref{supptab:aligner}, we show the performances of the aligners in terms of number of alignments that the aligners generate given the assembly and the reads to align, run time (wall clock), and the memory requirement.

\begin{table}[htb]
\begin{center}
\caption{\small Performance of the aligners}
\label{supptab:aligner}
\resizebox{\textwidth}{!}{
\begin{tabular}{|l|l|l|l|rrr|}
\hline Data Set for & Assembler & Aligner & Platform of the & Number of & Run Time & Memory\\
the Assembly & & & Aligned Reads & Alignments & & (GB) \\\hline
E.Coli K-12 - ONT & Miniasm & Minimap2 & ONT & 1,464,840 & 2m 08s & 1.77 \\
E.Coli K-12 - ONT & Canu & Minimap2 & ONT & 1,662,418 & 1m 41s & 1.76 \\
E.Coli K-12 - ONT(30X)& Canu & Minimap2 & ONT(30X) & 147,590 & 12s & 0.56 \\\hhline{|=|=|=|=|===|}
E.Coli O157 - PacBio & Miniasm & Minimap2 & PacBio & 739,715 & 1m 06s & 1.62 \\
E.Coli O157 - PacBio & Miniasm & Minimap2 & Illumina & 21,970,540 & 2m 38s & 3.00 \\
E.Coli O157 - PacBio & Miniasm & Minimap2 & PacBio + Illumina & 26,140,933 & 2m 58s & 3.22 \\
E.Coli O157 - PacBio & Canu & Minimap2 & PacBio & 741,626 & 1m 11s & 1.55 \\
E.Coli O157 - PacBio(30x)& Canu & Minimap2 & PacBio(30X) & 148,256 & 14s & 0.48 \\
E.Coli O157 - PacBio & Miniasm & BWA-MEM & Illumina & 19,661,998 & 7m 25s & 0.77 \\
E.Coli O157 - PacBio & Miniasm & BWA-MEM & PacBio + Illumina & 20,401,713 & 8m 12s & 0.79 \\
E.Coli O157 - PacBio & Canu & BWA-MEM & Illumina & 23,334,235 & 3m & 0.67 \\
E.Coli O157 - PacBio(30X)& Canu & BWA-MEM & Illumina & 23,326,199 & 3m 42s & 0.69 \\
E.Coli O157 - PacBio & Miniasm & pbalign & PacBio & 46,447 & 104h 10m 43s & 29.44 \\
E.Coli O157 - PacBio & Canu & pbalign & PacBio & 51,987 & 37m 03s & 61.10 \\\hhline{|=|=|=|=|===|}
CHM1 - PacBio & Miniasm & Minimap2 & PacBio & 2,140,172 & 20m 24s & 1.62 \\
CHM1 - PacBio & Canu & Minimap2 & PacBio & 3,309,401 & 8m 11s & 1.63 \\
CHM1 - PacBio & Miniasm & pbalign & PacBio & 98,985 & 98h 01m 55s & 21.30 \\
CHM1 - PacBio & Canu & pbalign & PacBio & 48,934 & 98h 01m 59s & 57.14\\\hline
\end{tabular}}
\end{center}
{\small We generate the assembly using the reads specified under \textit{Data Set for the Assembly}. We use Canu \cite{Koren2017} and Miniasm \cite{Li2016a} assemblers as specified in \textit{Assembler}. Here, the reads specified under \textit{Platform of the Aligned Reads} are aligned to the assembly using the \textit{Aligner}. We use Minimap2 \cite{Li2018} aligner for aligning both long and short reads to the assembly and BWA-MEM \cite{Li2009} aligner to align the short reads to the assembly. We report the performance of the aligners in terms of the number of the aligners (\textit{Number of Alignments}), the run time (\textit{Run Time}), and the maximum memory requirement \textit{Memory}.}
\end{table}

\clearpage
\section{Robustness of Apollo}

\begin{table}[htb]
\begin{center}
\caption{Apollo's robustness based on the chunk size of the long read and the contig}
\label{supptab:chunk}
\begin{tabular}{|l|l|rrr|}
\hline Long Read & Contig Chunk & Aligned & Coverage & Avg.\\
Chunk Size & Size & Bases & (\%) & Identity (\%) \\\hline
1000  & Original & 5,708,747 & 98.05 & 97.90 \\
1000  & 25000 & 5,487,736 & 94.46 & 97.33 \\
1000  & 50000 & 5,689,120 & 97.95 & 97.28 \\
1000  & 100000 & 5,493,663 & 94.52 & 97.27 \\
5000  & 25000 & 5,430,700 & 93.06 & 89.74 \\
5000  & 50000 & 5,411,163 & 92.68 & 89.71 \\
5000  & 100000 & 5,516,599 & 94.49 & 89.70 \\
10000  & 25000 & 5,415,333 & 92.65 & 89.18 \\
10000  & 50000 & 5,423,340 & 92.75 & 89.14 \\
10000  & 100000 & 5,474,159 & 93.61 & 89.14 \\\hline
\end{tabular}
\end{center}
{\small Here we divide the long reads and the assembly into smaller chunks. We use E.coli O157 data set where Miniasm generates the assembly. We divide long reads into smaller reads with lengths 1000, 5000, and 10000. Similarly, we divide the assembly contigs into smaller contigs with lengths 25000, 50000, and 100000. We align each chunked read to each chunked contig. We report the performance of Apollo given the chunked assembly and chunked reads.}
\end{table}

Here in Tables ~\ref{supptab:chunk}, ~\ref{supptab:deletion}, ~\ref{supptab:insertion}, ~\ref{supptab:transition}, we show the robustness of Apollo based on the parameters that has a direct affect on the machine learning algorithm. In each of the tables we show that Apollo is robust to different set of parameters.

\begin{table}[htb]
\begin{center}
\caption{Apollo's robustness based on the maximum deletion and filter size parameters}
\label{supptab:deletion}
\begin{tabular}{|l|l|rrr|}
\hline Max & Filter & Aligned & Coverage & Avg.\\
Deletion (-d) & Size (-f) & Bases & (\%) & Identity (\%) \\\hline
3  & 100 & 5,699,182 & 97.91 & 97.39 \\
5  & 100 & 5,696,138 & 97.93 & 97.35 \\
15  & 100 & 5,678,838 & 97.90 & 97.31 \\
3  & 200 & 5,705,130 & 98.12 & 97.51 \\
5  & 200 & 5,704,582 & 98.12 & 97.50 \\
15  & 200 & 5,702,478 & 98.14 & 97.51 \\\hline
\end{tabular}
\end{center}
{\small Performance of Apollo with respect to the parameter that defines the maximum number of deletion in one transition ($d=3$, $d=5$, $d=15$). We also adjust the filter size ($f=100$, $f=200$)}
\end{table}

\begin{table}[htb]
\begin{center}
\caption{Apollo's robustness based on the maximum insertion and filter size parameters}
\label{supptab:insertion}
\begin{tabular}{|l|l|rrr|}
\hline Max & Filter & Aligned & Coverage & Avg.\\
Insertion (-s) & Size (-f) & Bases & (\%) & Identity (\%) \\\hline
1  & 100 & 5,685,635 & 97.89 & 96.60 \\
5  & 100 & 5,638,585 & 97.62 & 96.96 \\
10  & 100 & 5,365,978 & 95.54 & 95.31 \\
1  & 200 & 5,685,040 & 98.02 & 96.68 \\
5  & 200 & 5,692,813 & 98.07 & 97.40 \\
10  & 200 & 5,623,736 & 97.62 & 97.01 \\\hline
\end{tabular}
\end{center}
{\small Performance of Apollo with respect to the parameter that defines the maximum number of insertion states for each basepair ($i=1$, $i=5$, $i=10$). We also adjust the filter size ($f=100$, $f=200$)}
\end{table}

\begin{table}[htb]
\begin{center}
\caption{Apollo's robustness based on the match transition, insertion transition probabilities, and the filter size parameters}
\label{supptab:transition}
\begin{tabular}{|l|l|l|rrr|}
\hline Match Transition & Insertion Transition & Filter & Aligned & Coverage & Avg.\\
Probability (-tm) & Probability (-ti) & Size (-f) & Bases & (\%) & Identity (\%) \\\hline
0.60  & 0.25 & 100 & 5,670,852 & 97.95 & 96.25 \\
0.60  & 0.30 & 100 & 5,660,957 & 97.90 & 95.96 \\
0.80  & 0.10 & 100 & 5,699,660 & 98.02 & 97.88 \\
0.90  & 0.05 & 100 & 5,685,770 & 97.89 & 97.74 \\
0.60  & 0.25 & 200 & 5,682,512 & 98.10 & 96.44 \\
0.60  & 0.30 & 200 & 5,681,993 & 98.13 & 96.18 \\
0.80  & 0.10 & 200 & 5,707,293 & 98.16 & 98.03 \\
0.90  & 0.05 & 200 & 5,695,902 & 98.05 & 97.89 \\\hline
\end{tabular}
\end{center}
{\small Performance of Apollo with respect to the parameters that define the match and insertion transition probabilities ($tm=60$ - $ti=0.25$, $tm=60$ - $ti=0.30$, $tm=80$ - $ti=0.10$, $tm=90$ - $ti=0.05$). We also adjust the filter size ($f=100$, $f=200$)}
\end{table}

\clearpage

\section{Parameters}

We show the parameter settings of the aligners that we used to align the reads to the assembly in Table ~\ref{supptab:aligner-params}.

\begin{table}[htb]
\begin{center}
\caption{List of the parameters that are used to align the reads to the assemblies}
\label{supptab:aligner-params}
\begin{tabular}{|l|c|}
\hline Aligner &  Parameters \\\hline
BWA-MEM & -t 10\\\hline
Minimap2 (for PacBio) & -x map-pb -a -k15 -w5 -t 10 -p 0.6\\\hline
Minimap2 (for ONT) & -x map-ont -a -k15 -w5 -t 10 -p 0.6\\\hline
pbalign & --nproc 10 \\\hline
\end{tabular}
\end{center}
\end{table}

\clearpage

\bibliographystyle{unsrt}
\bibliography{document}